\documentclass[10pt]{article}

\usepackage{amsmath}
\usepackage{amssymb}
\usepackage{ulem}
\usepackage{graphicx}
\usepackage{tabu}
\usepackage{cite}

\usepackage{color}

\usepackage{booktabs}
\usepackage[version=3]{mhchem}

\usepackage[labelfont=bf,labelsep=period,justification=raggedright]{caption}

\makeatletter
\renewcommand{\@biblabel}[1]{\quad#1.}
\makeatother

\date{}

\begin{document}

\begin{flushleft}
{\Large \textbf{Landscape and flux for quantifying global stability and dynamics of game theory}}
\\
Li Xu$^{1}$, Jin Wang$^{1,2,\ast}$
\\
\textbf{1} State Key Laboratory of Electroanalytical
Chemistry, Changchun Institute of Applied Chemistry, Chinese
Academy of Sciences, Changchun, Jilin, China
\\
\textbf{2} Department of Chemistry and Physics, State University of New York at Stony Brook, Stony Brook, NY, USA
\\
$\ast$ E-mail: jin.wang.1@stonybrook.edu
\end{flushleft}

\section*{Abstract}

 Game theory has been widely applied to many areas including economics, biology and social sciences. However, it is still challenging to quantify the global stability and global dynamics of the game theory.  We developed a landscape and flux framework to quantify the global stability and global dynamics of the game theory. As an example, we investigated the models of three-strategy games: a special replicator-mutator game, the repeated prison dilemma model. In this model, one stable state, two stable states and limit cycle can emerge under different parameters. The repeated Prisoner's Dilemma system has Hopf bifurcation transitions from one stable state to limit cycle state, and then to another one stable state or two stable states, or vice versa. We explored the global stability of the repeated Prisoner's Dilemma system and the kinetic paths between the basins of attractor. The paths are irreversible due to the non-zero flux. One can explain the game for $Peace$ and $War$.

\section*{Introduction}


Game theory is the study of conflict and cooperative strategic decision making between intelligent rational decision-makers. Game theory has widely been recognized to be important and useful in many fields such as economics, political science, psychology, computer science, biology etc. The game dynamics usually can converge to stable point attractors\cite{Hofbauer2011PSAM,Sandholm2009Encyclopedia}. However, a more complex non-equilibrium dynamics can lead to stable oscillations. The cyclical oscillations have been explored in the game-theoretical models of price dispersion\cite{Cason2003JET} and in the side-blotched lizard with three color morphs of males signalling three different alternative reproductive strategies(Uta stansburiana) experimentally\cite{Sinervo1996Nature}. Understanding the dynamical and global stability of the game theory is one of the greatest challenges at present. Even though numerous studies have been explored over the past decades with great advances in this field, there are still several unsolved problems.

The evolutionary stability was first introduced and formulated by Maynard Smith and Price in 1973\cite{Maynard1973Nature,Nowak2006Book}. They first applied the game theoretical ideas to evolutionary biology and population dynamics. This is the birth of evolutionary game theory which studies the behaviors of the large populations\cite{Sandholm2009Encyclopedia}. Evolutionary population game is a general framework for exploring the strategic interactions among large populations of agents. The agents play pure strategies and are with random matching. The dynamics of biological structure and population behaviors often has high stability and robustness. Thus, one of the central problem is to explore the stability and robustness of the evolutionary game theory in a population. The significant applications of game theory are on modeling and analyzing varieties of human and animal behaviors we observed around us. The game theory systems are complex and involve different interactions among the agents. The strategic interactions can lead to complex dynamics.

There have been many studies on the stability of game theory. However, most of the investigations are focused on the local stability, which are the analyses to uncover whether a system will restore to equilibrium under a small disturbance.  An evolutionary stable strategy (ESS) is a strategy or mixed strategy that all members play, and is resistent to invasions by a few mutants playing a different strategy in a population. But the system can move far from its ESS equilibrium since the it is under continuous small perturbations from the mutations and events by chance \cite{Foster1990TPB}. Thus, the ESS can not guarantee the global stability of the system. A Nash equilibrium (NE) is a result in a non-cooperative game with each player assumed to know the other players' equilibrium strategy and the agents get nothing by altering only their own strategy. The NE can be stable or unstable. It is very similar to an ESS\cite{Nowak2006Book} . The famous evolutionary stable strategy (ESS) and Nash equilibrium (NE) are insufficient conditions of dynamic stability since they are only local criterion under small fuctuations\cite{Foster1990TPB}. It is often not clear whether the system will reach equilibrium from an arbitrary initial state or whether the system can be switched from one locally stable state to another. These dynamical issues depend on the global structure of the system. Furthermore,  the link between the global characterization and the dynamics of the game theory systems is often missing. The global stability of the game theory is thus still challenging at present.

Deterministic population dynamics can only describe the average dynamics of the system. Both external and intrinsic fluctuations are unavoidable\cite{Swain2002PNAS}. The environmental fluctuations can influence the behaviors of population while the intrinsic fluctuations originated from mutations or random errors in implementations can not be neglected in finite population. They may play an essential role in the dynamics of the system. The analysis of stochastic evolutionary game dynamics was first studied by Foster and Young in 1990\cite{Foster1990TPB}. They defined the stochastic stability and the stochastically stable set which is a set of stochastically stable equilibrium states when the fluctuations tend to zero slowly in the long run(so called long-run equilibria)\cite{Foster1990TPB,Szabo2007PR}. The stable state sets are obtained by the potential theory\cite{Foster1990TPB}. However, the general approach for exploring the global stability of the game theory systems are still missing.

The researchers have also explored the game theory system with
the method of Lyapunov function which can be used to study the global stability. Some analytical Lyapunov functions were found for certain highly simplified game theory models\cite{Sandholm2009Encyclopedia}. However, it is still challenging to find the Lyapunov function of the game theory models with complex dynamics. In this study, we will provide a general approach to investigate the Lyapunov function. We will also develop a general framework for exploring the robustness and global stability of the game theory systems.

Besides uncovering the dynamics of the simple convergent stable states, exploring the mechanism of the non-convergent behavior is even more important for understanding the nature of the dynamics for evolutionary game theory. This is because certain more complicated behaviors such as oscillations and chaos  often emerge in human interaction\cite{Hommes2012GEB}. The most well known model of evolutionary dynamics is the replicator model. The simplest replicator dynamics of three-strategy games can only give certain behaviors: sinks, sources and saddles\cite{Hofbauer1980SJAM,Hommes2012GEB} or heteroclinic cycles for Rock-Paper-Scissors(RPS) game. However, the replicator dynamics can not provide a stable limit cycle behavior\cite{Zeeman1980Springer,Hommes2012GEB}. Lyapunov stable equilibria of the replicator dynamics are Nash equilibria and ESS of the replicator dynamics are asymptotically stable \cite{Pais2012SJADS,Weibull1995Book}. Mutation effects can be included in order to promote the chances that players change from one strategy to another spontaneously. The selection and mutation model has been explored in population genetics for decades. The replicator-mutator dynamics played a key role in evolutionary theory\cite{Pais2012SJADS,Weibull1995Book,Bladon,Allen,Nowak2006Book}.

In this study, we will develop a landscape and flux framework to quantify the global stability and robustness of evolutionary game theory. Conventional stability analysis of game theory (Nash equilibrium and ESS) give a static view and local description. In this study, we give a dynamical view and global quantification of the stability. We found the flux is additional force not present in the conventional game theory in determining the global dynamics. Both landscape and flux determines the dynamics. The landscape topography and kinetic transitions as well as optimal kinetic paths from one basin to another (local stable strategy) provide the global quantification of the strategy switching process and functional stability. That is not present in the current game theory. The global stability can be systematically studied in the landscape and flux approach via Lyapunov functions while these are only found for special cases (one dimensional case) in the current game theory. In the current evolutional game theory, the driving forces has never been decomposed. The landscape and flux theory provides a framework to quantify each component of the driving force and dictate the evolutionary game dynamics. We also explored non-equilibrium thermodynamics which is not covered in the current repeated Prisoner's Dilemma game or evolutionary game theory.

\begin{figure*}[!ht]
\centering
\includegraphics[width=0.7\textwidth]{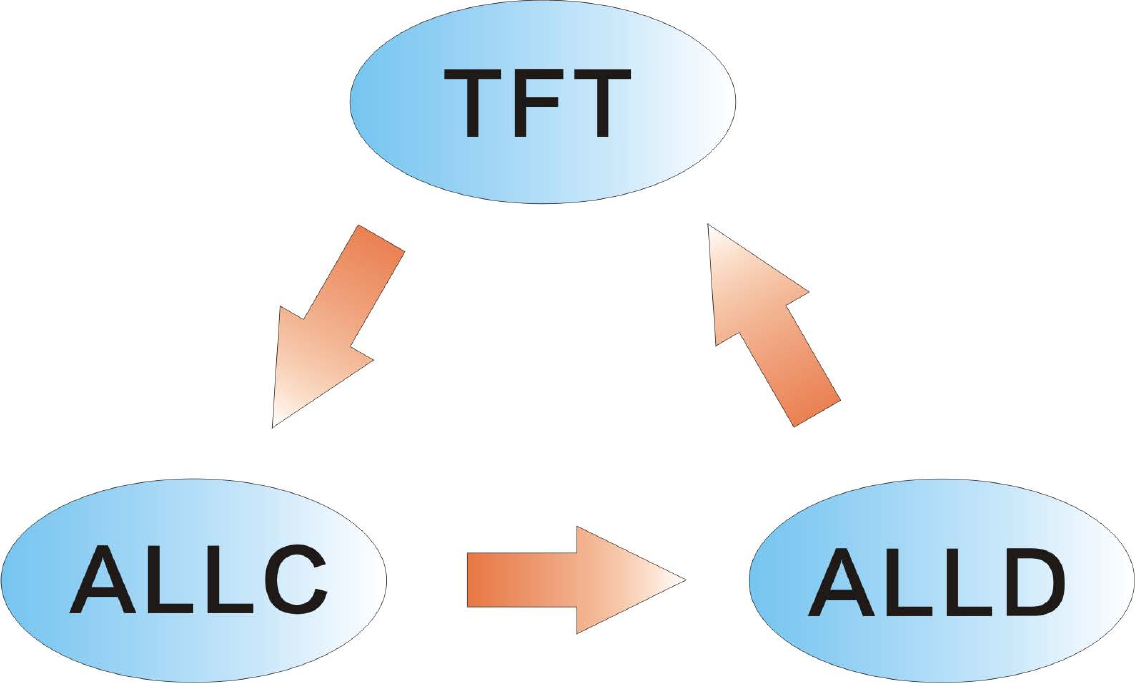}
\caption{The schematic of repeated prisoner's dilemma. $ALLC$ is short for always cooperate. $ALLD$ is short for always defect. $TFT$ is short for tit-for-tat.}\label{PD_schematic}%
\end{figure*}

We use a replicator-mutator model of the repeated Prisoner's Dilemma\cite{Imhof2005Evolutionary,Toupo2014IJBC,Pais,Pais2012SJADS} as an example to illustrate our general theory. There are three interaction strategies in this model: always cooperate simplified by $ALLC$, always defect simplified by $ALLD$, and tit-for-tat simplified by $TFT$. Figure \ref{PD_schematic} shows the schematic of repeated Prisoner's Dilemma. $ALLD$ players are the first winners for random initial strategies in the population. Then small numbers of $ALLC$ players will invade and replace strategy $ALLD$. The consideration of mutation effect can lead to the repeated Prisoner's Dilemma to a stable limit cycle state rather than a $ALLD$ dominant state since $ALLD$ is the only strict Nash equilibrium. The repeated Prisoner's Dilemma existed sustained oscillations among $ALLD$, $ALLC$ and $TFT$ strategies. We developed a landscape and flux landscape theory\cite{Wang2008PNAS,Wang2010JCP,Wang2010PNAS,Xu2013Non,Zhang2012JCP,Xu2014PLOSONE,Xu2014PLOSONE2} to explore the global behavior and dynamics of the evolutionary game theory of the Prisoner's Dilemma system. We quantified the population landscape related to the steady state probability distribution which determine the global behavior. The landscape has the shape of attractor basins for multi-stability and Mexican hat for limit cycle oscillations.  We also found that the intrinsic landscape with a Lyapunov function property for the the repeated Prisoner's Dilemma game model  can quantify the global stability. The non-equilibrium evolutionary game theory dynamics for repeated Prisoner's Dilemma is found to be determined by both the landscape and the curl flux. The curl probability flux can lead to the break down of the detailed balance and drive the stable periodical oscillation flow along the limit cycle ring\cite{Wang2008PNAS}. We explored the stability and robustness of repeated Prisoner's Dilemma game against the mutation rate and the pay-off matrix. The pathways between basins are quantified by the path integral method and irreversible due to the non-zero flux.

\section*{Models and Results}

\subsection*{The landscape and flux quantification for game theory}

The landscape and flux theory as well as non-equilibrium thermodynamics for the general dynamical systems have been explored in several different fields\cite{Wang2008PNAS,Wang2010JCP,Wang2010PNAS,Xu2013Non,Zhang2012JCP,Xu2014PLOSONE,Xu2014PLOSONE2, Wang2015AP}. They can be used to address the issues of global stability, function, robustness for dynamical systems. Here, we will apply the landscape and flux theory to quantify the global stability and robustness of the game theory.

Due to intrinsic and extrinsic fluctuations\cite{Swain2002PNAS}, the deterministic dynamics described by a set of ordinary differential equations are supplemented with the additional fluctuation force. Then the stochastic dynamics emerges \cite{VanKampen,Gillespie1977JPC}: $d \mathbf{x} = \mathbf{F}(\mathbf{x})dt + \mathbf{g} \cdot d \mathbf{W} $, where $\mathbf{x}$ is the state vector representing the population or species density in game dynamics), $\mathbf{F}(\mathbf{x})$ is the driving force, $\mathbf{W}$ coupled through the matrix $\mathbf{g}$ represents an independent Wiener process. The evolution of the stochastic dynamics is thus more appropriately described by probability evolution. The probability distribution $P(\mathbf{x},t)$ can be obtained by solving the corresponding Fokker-Planck equation\cite{VanKampen,Gillespie1977JPC}:
\begin{equation}\label{FPE}
\partial{P} / \partial{t} = -\nabla\cdot\mathbf{J} = -\nabla\cdot[\mathbf{F}P - (1/2) \nabla \cdot ( (\mathbf{g}\cdot\mathbf{g}^{\mathbf{T}}) P )] .
\end{equation}
We set $D \mathbf{D}=(1/2) (\mathbf{g} \cdot \mathbf{g}^{\mathbf{T}})$, where $D$ is a constant describing the scale of the fluctuations and $\mathbf{D}$ represents the anisotropy diffusion matrix of the fluctuations.

The game process can be treated as a binomial sampling process: $N$ players play the games as a sample of $2N$ players with different strategy from a large population of players.
So we can set $D_{ij} = x_i (\delta_{ij} - x_j)$ coming from the sampling nature of the game which is widely used in evolutionary population dynamics\cite{Zhang2012JCP}.

The matrix $D_{ij}$ has some special properties. The first is
\begin{equation}
(\nabla\cdot\mathbf{D})_i = 1 - n x_i ,
\end{equation}
so that $\sum\nolimits_{i=1}^{n} (\nabla\cdot\mathbf{D})_i = 0$.
The second is its inverse matrix is known to have the property \cite{Zhang2012JCP}:
\begin{equation}
(\mathbf{D}^{-1} \cdot \mathbf{F})_i = {F_i}/{x_i} - {F_n}/{x_n} ,
\end{equation}
where $F_n = - \sum\nolimits_{i=1}^{n-1} F_i$.

The steady state probability distribution $P_{ss}$ can be derived from the long time limit of the Fokker-Planck equation ${\partial{P}}/{\partial{t}}=0$. The steady state probability flux is defined as $ \mathbf{J_{ss}} = \mathbf{F}P_{ss} - D\nabla\cdot(\mathbf{D}P_{ss})$. The steady state flux is divergent free and therefore a rotational curl. The population landscape is defined as $U = - \mathrm{ln} P_{ss}$. The deterministic driving force $\mathbf{F}$ can be decomposed as: $\mathbf{F} = -D\mathbf{D}\cdot\nabla U + \mathbf{J}_{ss}/P_{ss} + D\nabla\cdot\mathbf{D}$. The flux $\mathbf{J}_{ss} = 0$ denotes the equilibrium with detailed balance while the flux $\mathbf{J}_{ss} \neq 0$ denotes the non-equilibrium with non-zero flux breaking the detailed balance. The dynamics of the equilibrium system is determined only by the population landscape gradient. The dynamics of the non-equilibrium system is determined by both the potential landscape and non-zero flux. \cite{Wang2008PNAS}

The intrinsic landscape at the zero fluctuation limit has the the property of Lyapunov function\cite{Graham1989,Sasai2003PNAS,Haken1987, Zhang2012JCP} and can  be used to quantify the global stability and function of the game theory systems. The intrinsic landscape $\phi_0$ follows the Hamilton - Jacobi equation as below\cite{Zhang2012JCP,Xu2013Non,Hu1986}:
\begin{eqnarray}\label{HJE}
\mathbf{F} \cdot \nabla \phi_0 + \nabla \phi_0 \cdot \mathbf{D} \cdot \nabla \phi_0 = 0.
\end{eqnarray}
and
\begin{equation}
\frac{d\phi}{dt} = \mathbf{F} \cdot \nabla \phi_0 = -\nabla \phi_0 \cdot \mathbf{D} \cdot \nabla \phi_0 \leq 0
\end{equation}
The intrinsic landscape $\phi_0$ is a Lyapunov function monotonously decreasing along a deterministic path\cite{Hu1986, Zhang2012JCP}. The intrinsic landscape $\phi_0$ quantifies the global stability for general dynamical systems either with or without detailed balance and can be solved by the level set method\cite{Mitchell2008JSC}.

The steady state probability distribution $P_{ss}$ and the population landscape $U$ have the relationship of $P_{ss}(\mathbf{N})=
\mathrm{exp}(-U)/Z$. $Z$ is the partition function $Z= \int
\mathrm{exp}(-U)d\mathbf{N}$. The entropy of the
non-equilibrium game system is given as
\cite{Wang2006PLOS,Zhang2012JCP,Ao2005PLR,Qian2009ME,Schnakenberg1976RMP}: $S=
- \int P(\mathbf{N},t) \mathrm{ln}P(\mathbf{N},t)d\mathbf{N}$, and
the energy is given as: $E= D\int
U P(\mathbf{N},t) d\mathbf{N}=- D \int \mathrm{ln} [Z P_{ss}]
P(\mathbf{N},t) d\mathbf{N}$. The nonequilibrium free energy
$\mathcal{F}$ is thus:
\begin{eqnarray}
\mathcal{F}= E -D S=D (\int P \, \mathrm{ln}(P/P_{ss}) \, d\mathbf{N} -
\mathrm{ln}Z).
\end{eqnarray}
The nonequilibrium free energy as the combination of energy and entropy reflects the first law of non-equilibrium
thermodynamics. The free energy decreases in time
monotonically while reaching its minimum value, $\mathcal{F}= -D
\mathrm{ln}Z$\cite{Wang2006PLOS,Zhang2012JCP,Ao2005PLR,Qian2009ME,Schnakenberg1976RMP}. This reflects the second law of non-equilibrium
thermodynamics. The free energy as a Lypunov functional can be used to explore the global
stability of stochastic non-equilibrium systems with finite fluctuations.

\subsection*{The repeated Prisoner's Dilemma game theory model with mutations}

We consider a finite set of pure strategies for a large finite population of players with random matching. Each player chooses a pure strategy from the strategy set $S={1,2,3,...,n}$\cite{Hofbauer2011PSAM,Sandholm2009Encyclopedia}. The aggregated behaviors of these players are described by a population state $\bf x$, with $x_i$ representing the proportion of players choosing pure strategy $S_i$. And $x_i$ represents the frequency of strategy $S_i$. $\bf A$ is the payoff matrix. The scalar $A_i(x)$ represents the payoff to strategy $S_i$ when the population state is $x$. Since the sum of all frequencies equals to 1: $\sum_i{x_i} = 1$, the system becomes $n-1$ dimensional. The average fitness (pay-off) $\bar f$ is obtained as $\bar f=\bf{xAx}$ by the members of the population. The fitness denotes the individual's evolutionary success \cite{Szabo2007PR}. In the game theory, the payoff of the game is the fitness. The fitness to strategy $i$ becomes $f_i=(\bf{Ax})_i$\cite{Hofbauer2011PSAM,Sandholm2009Encyclopedia}. In this study, we use simple three-strategy game which can be reduced to two dimensional. Here we will introduce the replicator-mutator dynamics.

The players in standard Prisoner's Dilemma model play either cooperate strategy or defect strategy simultaneously\cite{Rapoport}. The players earn their payoff depending on their choices of strategies. The mutually aided cooperators will acquire the reward $R$ when the cooperators are encountered. The defectors will obtain a punishment $Pu$ when the defectors are encountered. A cooperator acquires a sucker payoff $S$ and a defector acquires a temptation payoff $T$ when they encountered\cite{Toupo2014IJBC,DaniellePRE}. Prisoner's Dilemma should satisfy the relationship $T > R > Pu > S$. Mutual cooperative strategy is better than mutual defective strategy, so the reward $R$ should be greater than the punishment $Pu$. Defectors gain more temptation $T$ than the reward $R$ that cooperators gain if the partner cooperates. The defectors will loose less punishment $Pu$ than the sucker $S$ lost by cooperators if the partner defects.

We explored a repeated Prisoner's Dilemma model with mutation using replicator dynamics. Replicator dynamics was introduced by Taylor and Jonker (1978)\cite{Taylor1978MB}, which is the best-known dynamics in the models of biological evolution. The mean dynamic evolutionary equation is replicator dynamics shown as below\cite{Hofbauer2011PSAM,Sandholm2009Encyclopedia}

\begin{equation}
\frac{dx_i}{dt}=x_i(f_i- \bar f)=x_i((\bf{Ax})_i- \bf{xAx})
\end{equation}

In the presence of mutations, $ALLD$ and $ALLC$ players may all mutate to strategy $TFT$. Therefore, more players choose $TFT$ strategy, and will help $ALLC$ players to win the game. Then the $ALLD$ will emerge again to obtain more profit\cite{Nowak2006Book}. This dynamics can be used to quantify the peace and war oscillation\cite{Nowak2006Book}.

We define the probability that agents with strategy $S_i$ mutating to strategy $S_j$ as $q_{ij}$, which satisfies $\sum_{j}{q_{ij}}=1$. Thus, mutation matrix is $Q = [q_{ij}]$\cite{Pais,Nowak2006Book,Imhof2005Evolutionary,Pais2012SJADS}. The elements of the mutation matrix $Q$ are defined in terms of a mutation parameter $\mu$ satisfying $0 \leq \mu \leq 1 $. The mutation $\mu$ denotes the error probability in the process of replication. $\mu = 0$ denotes no mutation with perfect replication while $\mu=1$ denotes the mutation entirely. The replicator-mutator dynamics describes the dynamics of the population distribution
$\bf{x}$ as a result of replication driven by fitness $\bf{f}$ and mutation driven by $Q$\cite{Bladon,Allen,Nowak2006Book}:

\begin{eqnarray}\label{replicator-mutator_equation}
\frac{dx_i}{dt}=\sum_{j=1}^{N}{x_if_i\bf(x)q_{ji}}- x_i\bar f
\end{eqnarray}

This equation is the quasispecies equation proposed by Manfred Eigen and Peter Schuster \cite{Eigen}. We set a uniform probability of mutation from one strategy to another strategy with $q_{ii}=1-2\mu, q_{ij}=\mu$, then the matrix $Q$ is shown as follows\cite{Bladon,Allen,Nowak2006Book}:

\begin{equation}\label{Q}
Q=
\left(  \begin{array}{ccc}   
 1-2\mu & \mu & \mu\\  
 \mu & 1-2\mu  & \mu\\  
\mu & \mu  & 1-2\mu\\
\end{array}\right)
\end{equation}

The $x_1$,$x_2$,$x_3$ are the fractions of the population choosing the strategies $ALLD$,$TFT$,$ALLC$, respectively. We substitute $Q$ into Eq.\ref{replicator-mutator_equation}, then replicator-mutator dynamics are shown as the following simplified equations\cite{DaniellePRE,Toupo2014IJBC}:

\begin{eqnarray}\label{RPS_equation}
{dx_1}/{dt}=x_1(f_1-\bar{f})+\mu(-2 x_1 f_1+x_2 f_2+x_3 f_3) \nonumber\\
{dx_2}/{dt}=x_2(f_2-\bar{f})+\mu(-2 x_2 f_2+x_1 f_1+x_3 f_3) \nonumber\\
{dx_3}/{dt}=x_3(f_3-\bar{f})+\mu(-2 x_3 f_3+x_1 f_1+x_2 f_2)
\end{eqnarray}

We considered the players playing with infinite number of rounds. So in these limits, the average payoff matrix with cost are shown as\cite{Imhof2005Evolutionary,Toupo2014IJBC} for the strategies $ALLD$,$TFT$,$ALLC$:

\begin{equation}\label{A_Co}
A =
\left(  \begin{array}{ccc}   
 Pu & Pu & T\\  
 Pu-c & R-c  & R-c\\  
S & R  & R\\
\end{array}\right)
\end{equation}

$TFT$ strategy is conditional while $ALLD$ and $ALLC$ strategies are unconditional. Thus, the payoff value of strategy $TFT$ may have a small complexity cost\cite{Binmore1991Evolutionary,Imhof2005Evolutionary}.

The parameters are set as: $T=5$,$R=3$,$Pu=1$,$S=0$\cite{Toupo2014IJBC,Axelrod,Nowak2006Book}. $\mu$ is the average mutation rate between each two of three strategies. $c$ is a complexity cost for playing $TFT$. Larger $c$ represents the less players who play the strategy $TFT$.

\subsection*{Phase diagrams, Hopf bifurcations and limit cycles in evolutionary game dynamics upon $TFT$ cost changes}

\begin{figure*}[!ht]
\centering
\includegraphics[width=1.0\textwidth]{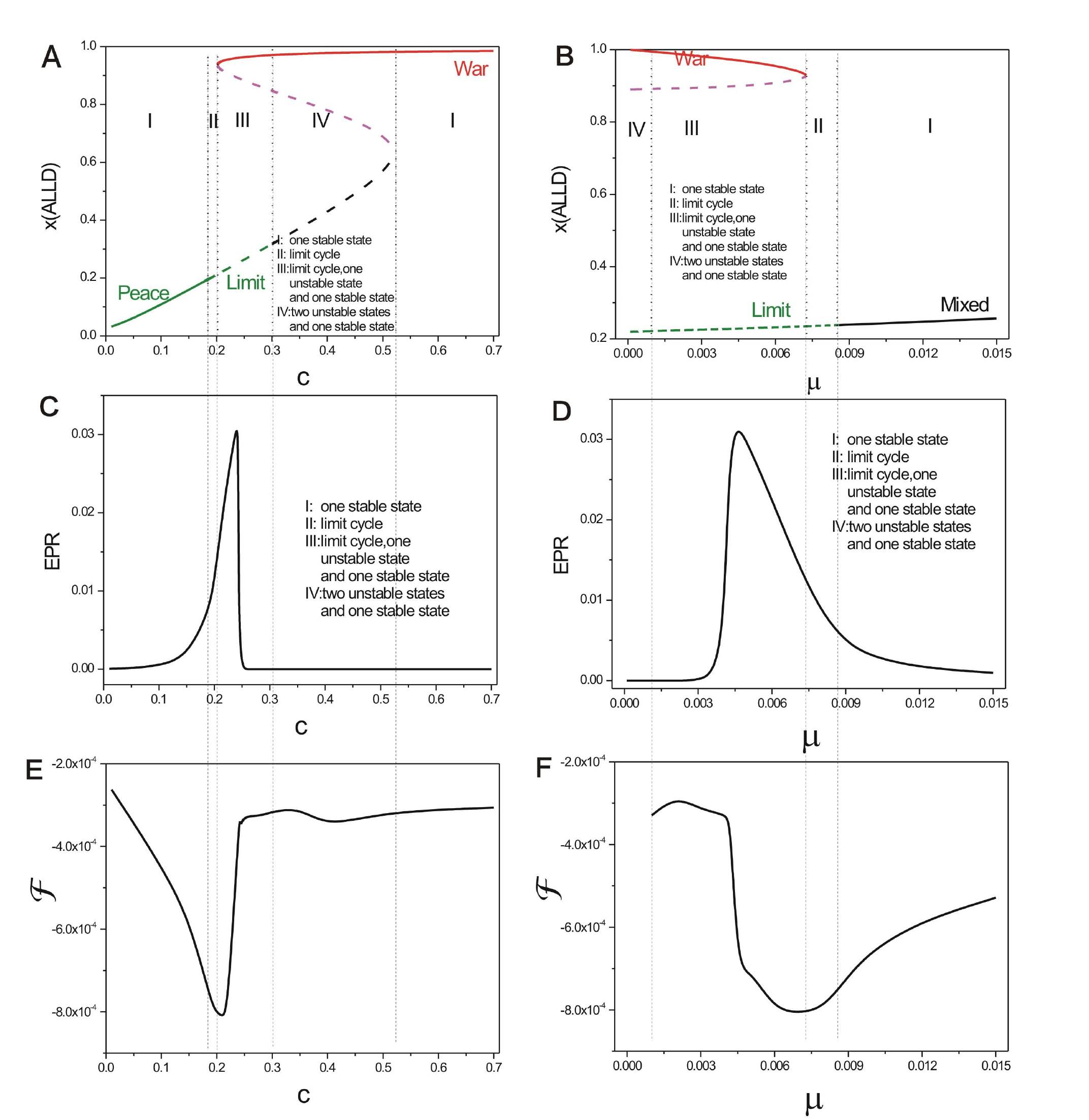}
\caption{A: the phase diagram for the repeated Prisoner's Dilemma with the constant parameter $\mu=0.006$ and changing cost parameter $c$. C: the entropy production rate versus parameter $c$. E: the free energy versus parameter $c$. B: the phase diagram for the repeated Prisoner's Dilemma with the constant parameter $c=0.22$ and changing parameter $\mu$. D: the entropy production rate versus parameter $\mu$. F: the free energy versus parameter $\mu$. The other parameters are $T=5$,$R=3$,$Pu=1$,$S=0$.}\label{cm_phase_EPR}%
\end{figure*}

Figure \ref{cm_phase_EPR} (A) shows the phase diagram which has a $S$ shape for the repeated Prisoner's Dilemma with the constant parameter $\mu=0.006$ and changing parameter $c$. There are five regions in this phase diagram. When the cost for $TFT$ $c$ is smaller, the game theory system has only one stable state, which denotes the $Peace$ state in the peace and war game shown in Region I. As $c$ increases, the stable state becomes a limit cycle in Region II. Then, as $c$ increases further, another new stable state (can be viewed as $War$ state) and an unstable saddle state emerge beside the limit cycle. It is a saddle-node bifurcation shown in Region III. As cost $c$ increases furthermore, the limit cycle diminishes and becomes an unstable state, along with a stable state ($War$ state) in Region IV. As cost $c$ increases even further, there is only one stable state ($War$ state) in Region V.

\subsubsection*{1. Population, intrinsic landscape and curl flux for quantifying global stability of the game theory dynamics}

\begin{figure*}[!ht]
\centering
\includegraphics[width=1.0\textwidth]{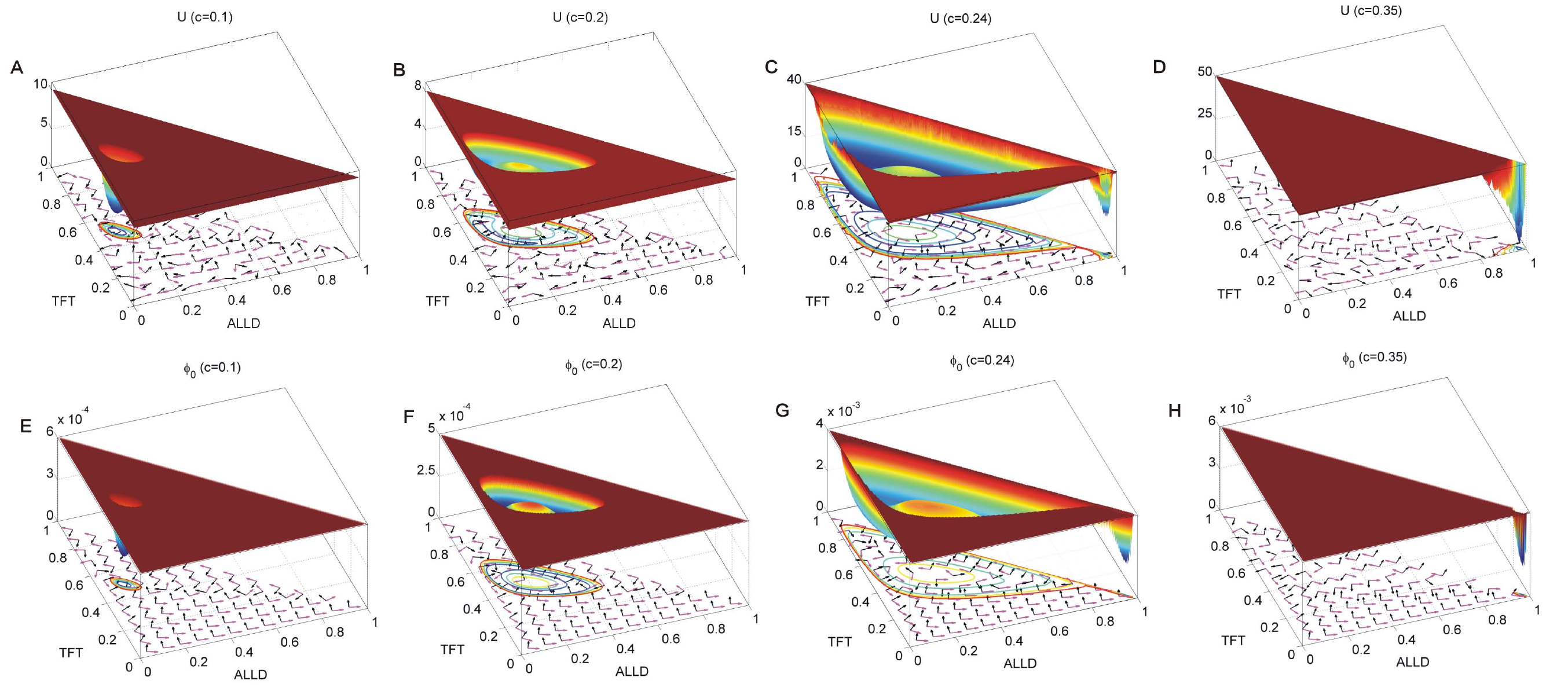}
\caption{The 3 dimensional population landscapes $U$ with increasing parameter $c$ are shown in $A,B,C,D$. Purple arrows represent the flux
velocity($\mathbf{J}_{ss}/P_{ss}$) while the black arrows
represent the negative gradient of population potential($-\nabla
U$). The 3 dimensional intrinsic energy landscapes $\phi_0$ with increasing parameter $c$ are shown in E,F,G,H. Purple arrows represent the intrinsic flux
velocity ($\mathbf{V}=(\mathbf{J}_{ss}/P_{ss})_{D \rightarrow
0}$) while the black arrows represent the negative gradient of
intrinsic potential($-\nabla \phi_0$).}\label{c_Uphi}%
\end{figure*}

By solving the Fokker-Planck diffusion equation, we obtain the steady distribution of the probability. Thus,  the population landscape of the system can be obtained as: $U=-lnP_{SS}$. We solved the Hamilton-Jacobi equation to obtain the
intrinsic landscape $\phi_0$ by the level set method\cite{Mitchell}.
Figure \ref{c_Uphi} show the 3 dimensional non-equilibrium population landscape $U$ at the top row and intrinsic potential landscape $\phi_0$ at the bottom row as the parameter $c$ increases, when the other parameters are set as $\mu=0.006$, $T=5$, $R=3$, $Pu=1$, $S=0$. We can see the underlying population landscape and intrinsic landscape have similar shapes in each column. The population landscape and the intrinsic landscape both have a basin shown in Figure \ref{c_Uphi}(A)(E) with small cost $c=0.1$. Figure \ref{c_Uphi}(B)shows the population landscape $U$ with a closed inhomogeneous ring valley which is not uniformly distributed, while Figure \ref{c_Uphi}(F) shows that the intrinsic landscape $\phi_0$ with Lyapunov properties has a closed homogeneous ring valley with cost $c=0.2$. The value of
$\phi_0$ along this ring valley is almost a constant. As cost $c=0.24$, a new basin emerged at the right corner of the space shown in Figure \ref{c_Uphi}(C)(G). It denotes the $War$ state with larger probability of strategy $ALLD$, which is a strict Nash equilibrium. Then at the cost $c=0.35$, the closed ring valley disappeared, only the stable $War$ state basin is left shown in Figure \ref{c_Uphi}(D)(H).

$-\nabla U$ is the negative gradient of the population landscape while $-\nabla \phi_0$ is the negative gradient of the intrinsic landscape. $-\nabla U$ at the top row and $-\nabla \phi_0$ at the bottom row are represented by black arrows. $\mathbf{J}_{ss}/P_{ss}$ is the steady state flux divided by steady state probability while $V=\mathbf{J}_{ss}/P_{ss})|_{D\rightarrow0}$ is the intrinsic flux velocity. $\mathbf{J}_{ss}/P_{ss}$ at the top row and $V$ at the bottom row are represented by purple arrows. We can see the flux with purple arrows are almost orthogonal to the negative gradient of $U$ under the black arrows around the basins or the closed ring valley shown on the top row. The region with higher population potentials has some disordered oriented arrows due to the lower probability and limit of computational accuracy. The flux velocity with purple arrows are orthogonal to the negative gradient of $\phi_0$ under black arrows at the bottom row. The landscape's gradient force $\nabla U$ or the $\nabla \phi_0$ attract the system to the basin or the closed ring valley, while the flux drives the periodical oscillation flow or spiral descent to the basin. It is necessary to characterize this non-equilibrium repeated Prisoner's Dilemma with both landscape and flux.

Game theory systems are often non-equilibrium open systems. They exchange energy from the environments which lead to dissipation. The entropy production rate and heat dissipation rate are equal in steady state. Thus, the entropy production rate is another global physical characterization of a non-equilibrium system and measured by the following formula:$EPR=\int \mathbf{J}^{2}/Pd\mathbf{x}$(\cite{Prigogine, Ge,Qian2006, Wang2008PNAS, Zhang2012JCP}). Figure \ref{cm_phase_EPR} (B) shows the entropy production rate ($EPR$) versus cost $c$. We can see $EPR$ has a bell shape in Region II and Region III with limit cycle in the system. It indicates that the limit cycle costs more energy to maintain.

\subsubsection*{2. Rational of formation of attractor state and landscape of game dynamics}

\begin{figure*}[!ht]
\centering
\includegraphics[width=1.0\textwidth]{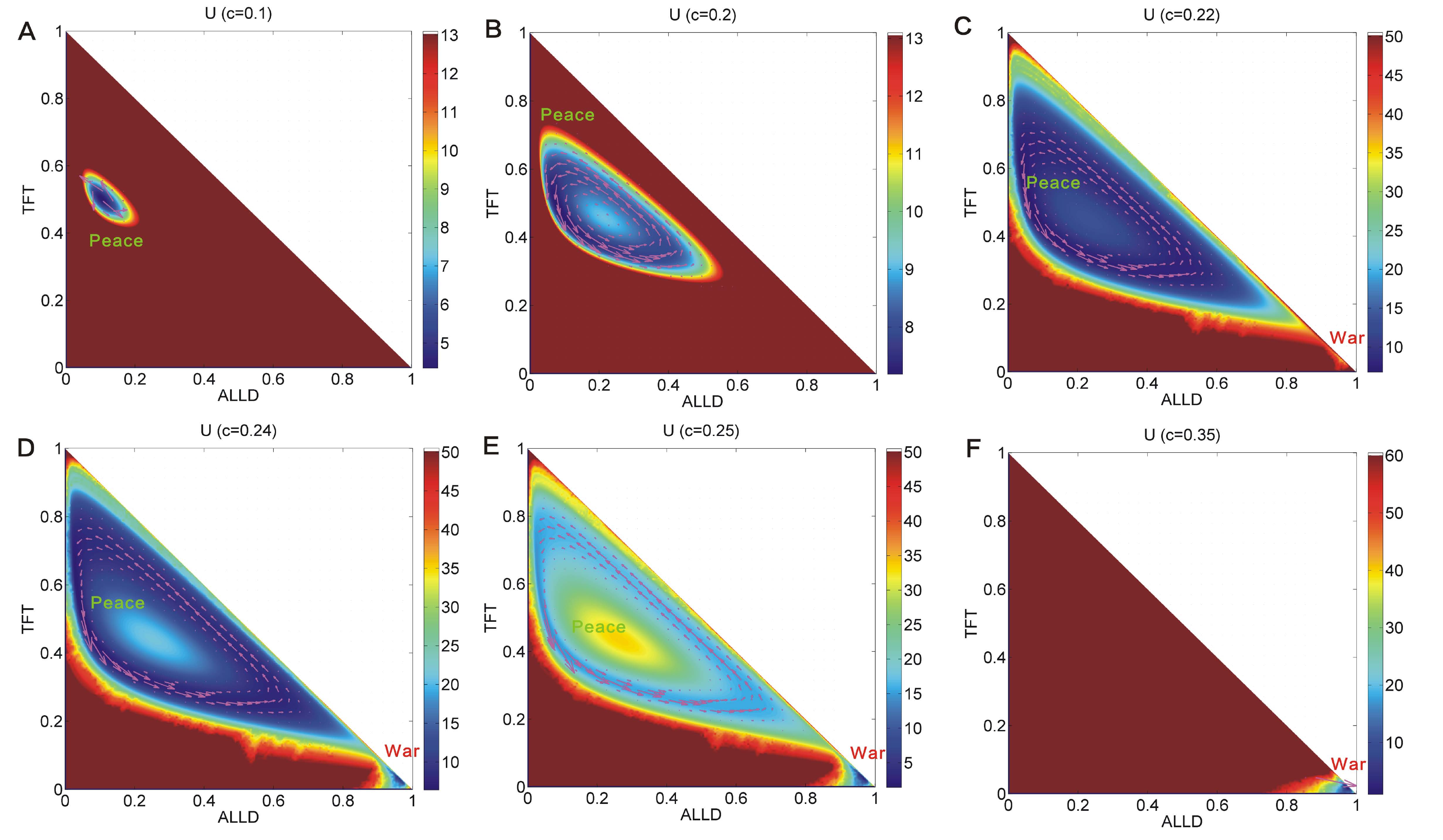}
\caption{The 2 dimensional population landscape $U$ with different parameter $c$ and constant parameter $\mu=0.006$. A:$c=0.1$, B:$c=0.2$, C:$c=0.22$, D:$c=0.24$, E:$c=0.25$, F:$c=0.35$. The purple arrows represent the flux.}\label{c_flux}%
\end{figure*}

We show the 2 dimensional population landscape $U$ for increasing parameter cost $c$ with constant $\mu=0.006$ in Figure \ref{c_flux}. Figure \ref{c_flux}(A) shows that the population landscape $U$ has a stable basin near the middle of $TFT$ axis with the cost parameter $c=0.1$. When the cost parameter for $TFT$ is small, the profit for $TFT$ players is more than the that with larger $c$. In this attractor, most players choose strategies of $ALLC$ and $TFT$. The fluxes represented by the purple arrows rotate anticlockwise around this stable state. This state can be viewed as "Peace" state in peace and war game. As cost parameter for $TFT$ increases $c=0.2$, a limit cycle emerges and replaces the stable state in Figure \ref{c_flux}(B). The population landscape has a blue ring valley along the deterministic trajectory. More players choose $TFT$ and $ALLC$ in the $Peace$ state. As more players mutate to $ALLC$, a small number of $ALLD$ players emerge. This leads to a state with more $ALLD$ players. As the $ALLD$ players become more, the profit obtained from the game becomes less.  Some $ALLD$ players convert their strategy to $TFT$. This makes a circle in the state space of strategy probability. Notice that the ring valley is not homogeneous in landscape depth. There is a deeper area on the left side of the limit cycle, which is still close to $TFT$ axis. This indicates that the $Peace$ state with deeper depth is more stable than other state. The system will stay in $Peace$ state much longer than any other state. Figure \ref{c_flux}(C) shows the ring valley of the oscillation expands its amplitude in the strategy-frequency space. A stable state $War$ emerges at the right corner of the triangle, which is close to the $ALLD \rightarrow 1$. The stable $War$ state is the one where most of players choose the $ALLD$ strategy. We can see that the limit cycle and the stable state coexist in the strategy-frequency space. It indicates that the system sometimes is in the limit cycle and sometimes in the stable state. System can switch between the $Peace$ and $War$ attractor basins states under the fluctuations and the mutations. When the cost for $TFT$ increases to $c=0.24$ and $c=0.25$ shown in Figure \ref{c_flux}(D)(E), the ring valley becomes shallower and shallower while the basin of the stable $War$ state becomes deeper and deeper. As $c$ increases, the profits obtained from the $TFT$ strategy decreases in the whole game, more players give up $TFT$ strategy and choose $ALLD$ strategy to earn more which lead to the more stable and deeper $War$ state basin. When the cost for $TFT$ increases to $c=0.35$ shown in Figure \ref{c_flux}(F), the oscillation ring valley disappears and the $War$ state survives and becomes deeper and more stable. Figure \ref{c_flux} also shows the flux (purple arrows) as mutation rate $\mu$ increases. We can see the fluxes has a anticlockwise curl nature along with limit cycle. The flux is the driving force for the stable oscillation flow.

\subsubsection*{3. Quantification of landscape topography of game dynamics}

\begin{figure*}[!ht]
\centering
\includegraphics[width=1.0\textwidth]{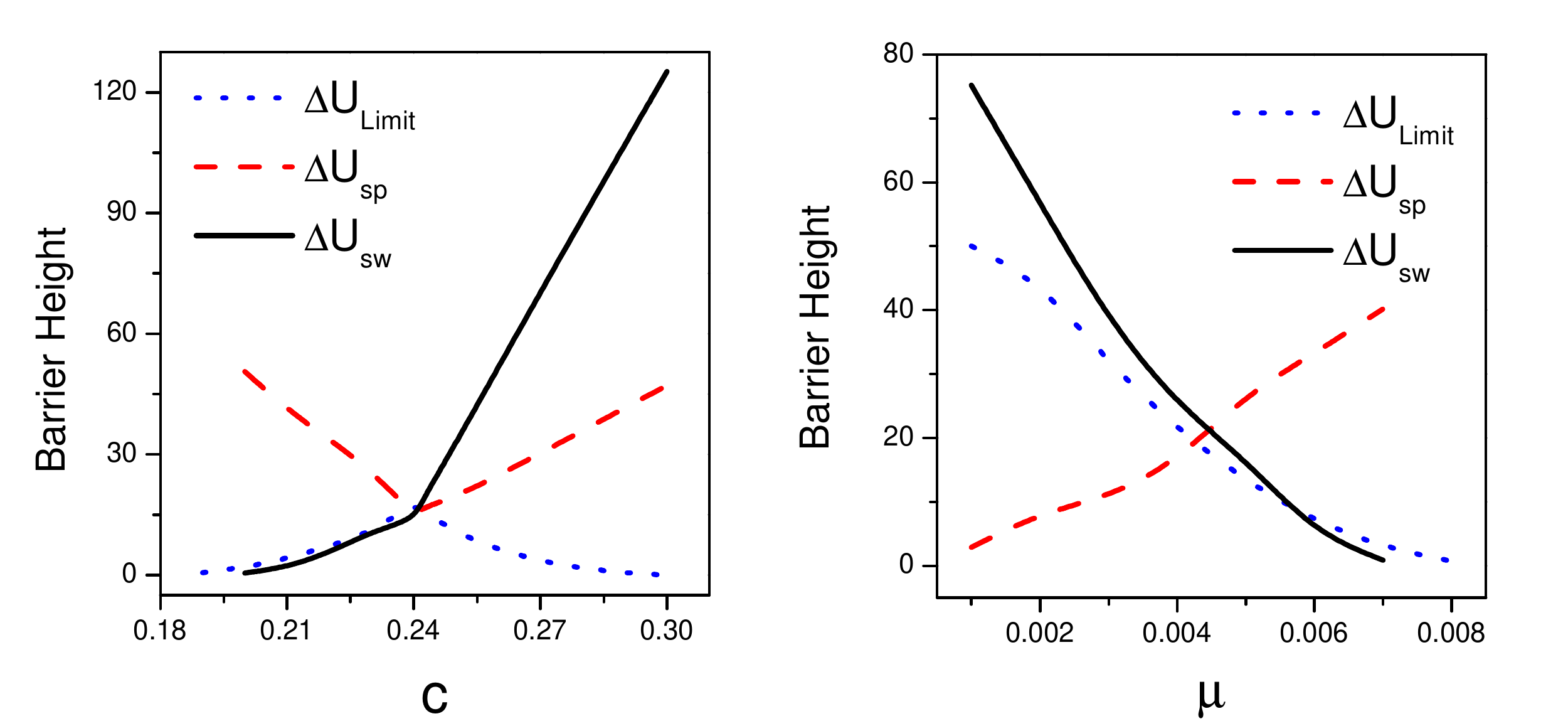}
\caption{A: The barrier heights versus parameter $c$ with parameter $\mu=0.006$. B: The barrier heights versus parameter $\mu$ with parameter $c=0.22$. Here, $\Delta U_{Limit}=U_o-U_p$, $\Delta U_{sp}=U_s-U_p$, $\Delta U_{sw}=U_s-U_w$. $U_o$ is the value of population potential landscape $U$ at the maximum point on the swell in the center of the limit cycle. $U_s$ is the value of population potential landscape $U$ at the saddle point between the limit cycle valley and the stable state basin $War$. $U_p$ is the minimum value of population potential landscape $U$ along the limit cycle near the $y$ axis, which is the $Peace$ state. $U_w$ is the minimum value of population potential landscape $U$ at the stable state $War$.}\label{cm_Barrier}%
\end{figure*}

We quantify the landscape topography and show barrier heights versus cost $c$ with mutation parameter $\mu=0.006$ in Figure \ref{cm_Barrier}(A). We first set $U_o$ as the value of population landscape $U$ at the maximum point at the center of the limit cycle. $U_s$ is the value of population landscape $U$ at the saddle point between the limit cycle valley and the stable $War$ state basin. $U_p$ is the minimum value of population landscape $U$ along the limit cycle near the $y$ axis, which is the $Peace$ state. $U_w$ is the minimum value of population landscape $U$ at the stable $War$ state. We set the barrier height for the oscillation ring valley as $\Delta U_{Limit}=U_{o}-U_{p}$, the barrier height between the saddle point and the oscillation as $\Delta U_{sp}=U_{s}-U_{p}$ and the barrier height between the saddle point and the $War$ stable state as $\Delta U_{sw}=U_{s}-U_{w}$. We can see as the cost $c$ increases, barrier height $\Delta U_{sw}$ increases,  barrier height $\Delta U_{sp}$ decreases first then increases, barrier height $\Delta U_{Limit}$ increases first then decreases. It indicates that the oscillation itself relative to the maximum point in the center of limit cycle becomes more stable first then becomes less stable. It has a turning point during the process of $c$ increasing. The $War$ state become more robust, and the barrier height from oscillation to $War$ state $\Delta U_{sp}$ becomes less than that of $\Delta U_{sw}$ for larger cost $c$ value. This implies that the $War$ attractor state becomes more preferred than that of oscillation, as the cost increases further.

\subsection*{Phase diagrams, state switching, landscapes and fluxes upon mutations}

Figure \ref{cm_phase_EPR}(B) shows the phase diagrams for the repeated Prisoner's Dilemma with changing parameter $\mu$ at the constant parameter $c=0.22$. There are four regions in phase diagram. When the mutation rate $\mu$ is smaller, the system has a stable state (can be viewed as $War$ state) in Region IV. As $\mu$ increases, a stable state $War$, an unstable saddle state and the limit cycle coexist in Region III. As the mutation rate $\mu$ increases further, the stable state and saddle point disappear after the saddle-node bifurcation. There is only limit cycle left in Region II. As the mutation rate $\mu$ keeps on increasing, the limit cycle becomes a stable $mixed$ state (with moderate probability of more than $20\%$ of the players with $ALLD$ strategy) in Region I. The $mixed$ state is the combination of these three strategies. We can see energy cost $EPR$ shown in Figure \ref{cm_phase_EPR}(D) has a bell shape as the mutation rate $\mu$ increases when the cost $c$ is moderate. This is because that the state of oscillation costs more energy in the strategy probability state space while one state cost less energy to maintain.

\begin{figure*}[!ht]
\centering
\includegraphics[width=1.0\textwidth]{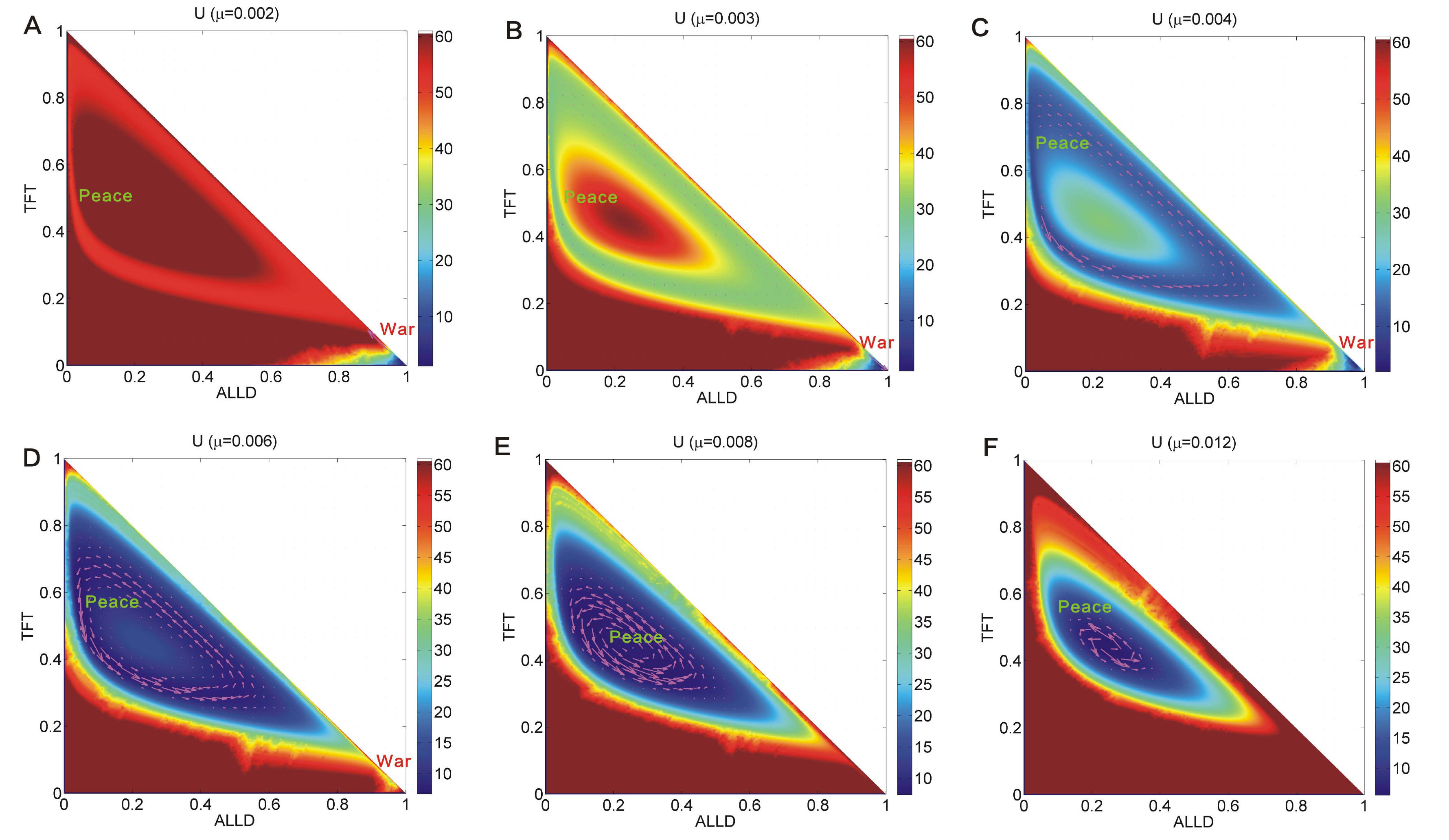}
\caption{The flux (purple arrows) and the gradient of the population landscape (black arrows) on the population potential landscape $U$ with different parameter $\mu$ and constant parameter $c=0.22$. A:$\mu=0.002$, B:$\mu=0.003$, C:$\mu=0.004$, D:$\mu=0.006$, E:$\mu=0.008$, F:$\mu=0.012$.}\label{m_flux}%
\end{figure*}

We can also see this process of transition in Figure \ref{m_flux} which shows that the population landscape $U$ and the flux change with the increasing mutation rate $\mu$ at constant cost $c=0.22$. Figure \ref{cm_Barrier}(B) shows that $\Delta U_{sp}$ increases and $\Delta U_{sw}$, $\Delta U_{Limit}$ decrease as mutation rate $\mu$ increases. This shows that when mutation rate is small, the limit cycle ring valley is very stable relative to its oscillation center, but has less probability relative to the $War$ state since the basin state $War$ is much deeper. It indicates that more players choose strategy $ALLD$, and the players do not like to mutate to the other two strategies. This leads to more stable $War$ state. As $\mu$ increases, the limit cycle ring valley becomes less stable relative to its oscillation center, but becomes more stable relative to the $War$ state. Eventually a much stable state $Peace$ emerges. The state $War$ becomes shallower and less stable, and finally diminishes as increasing $\mu$.

\subsection*{Phase diagrams, state switching, landscapes and fluxes upon temptation payoff}

\begin{figure*}[!ht]
\centering
\includegraphics[width=1.0\textwidth]{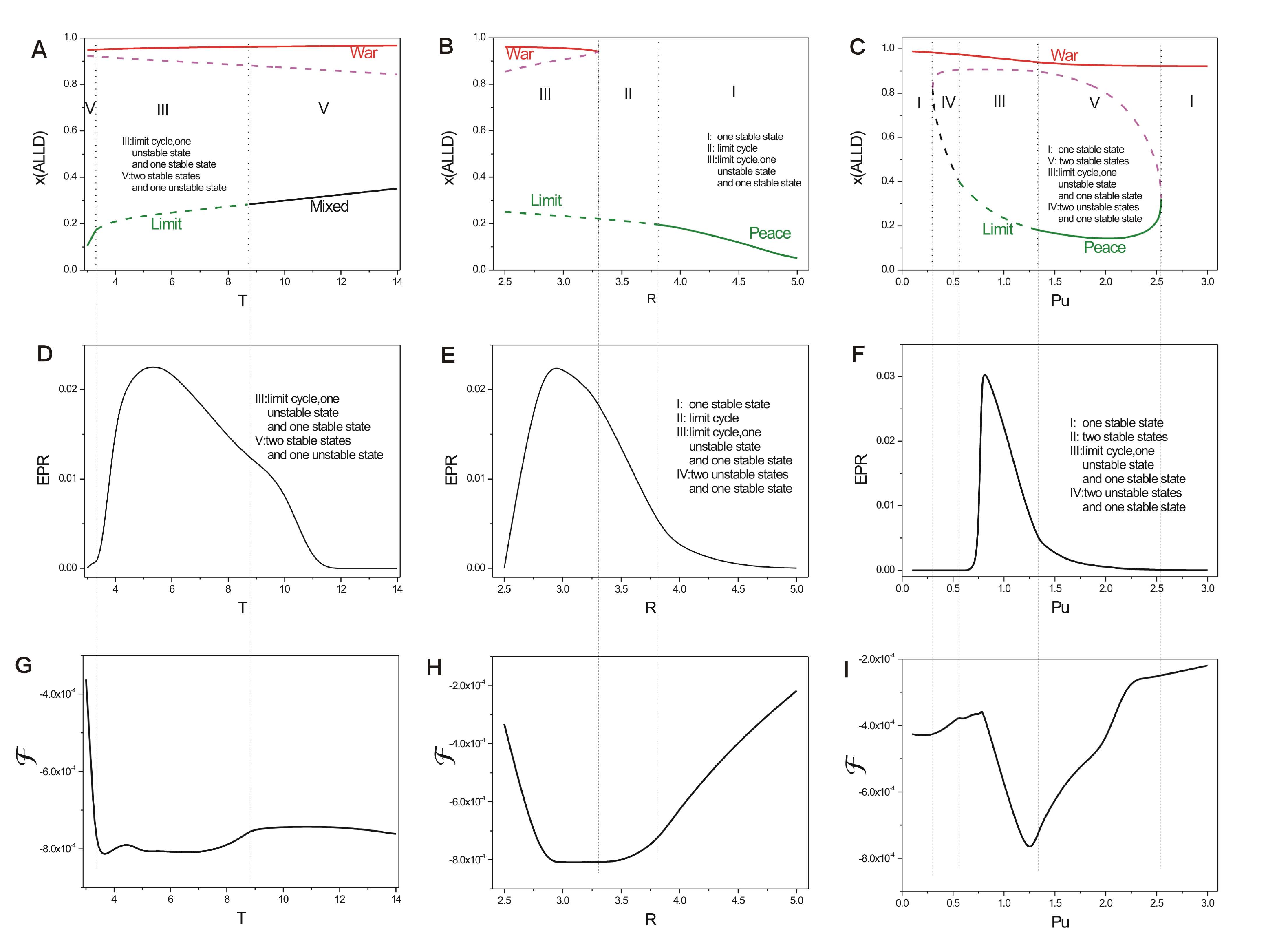}
\caption{The phase diagrams for the repeated Prisoner's Dilemma with changing parameter $T$ (A) $R$ (B), $Pu$ (C). The entropy production rate versus parameter $T$ (D), $R$ (E), $Pu$ (F). The free energies versus parameter $T$ (G), $R$ (H), $Pu$ (I). The other parameters are $S=0$,$\mu=0.006$,$c=0.22$}\label{TRP_phase_EPR}%
\end{figure*}

Figure \ref{TRP_phase_EPR}(A) shows the phase diagram for the repeated Prisoner's Dilemma with the constant parameters $\mu=0.006,c=0.22,R=3.0,Pu=1.0,S=0.0$ and changing parameter $T$.  $T$ is the temptation payoff earns by the defector while the cooperator gets sucker payoff $S$. There are three regions in this phase diagram. When the temptation $T$ is smaller, the game theory system has two stable states which can denote $Peace$ and $War$ in the peace and war game shown in Region left V.  As the temptation $T$ increases, a limit cycle emerges from the coexistence of $Peace$ and $War$ states in Region III.  As the temptation $T$ increases further, the limit cycle diminishes and becomes a stable $mixed$ state, along with a stable state ($War$ state) at the right of Region V. We can see $EPR$ shown in Figure \ref{TRP_phase_EPR}(D) also has a bell shape as the temptation $T$ increases. This also implies that limit cycle state cost more energy to maintain.

\begin{figure*}[!ht]
\centering
\includegraphics[width=0.8\textwidth]{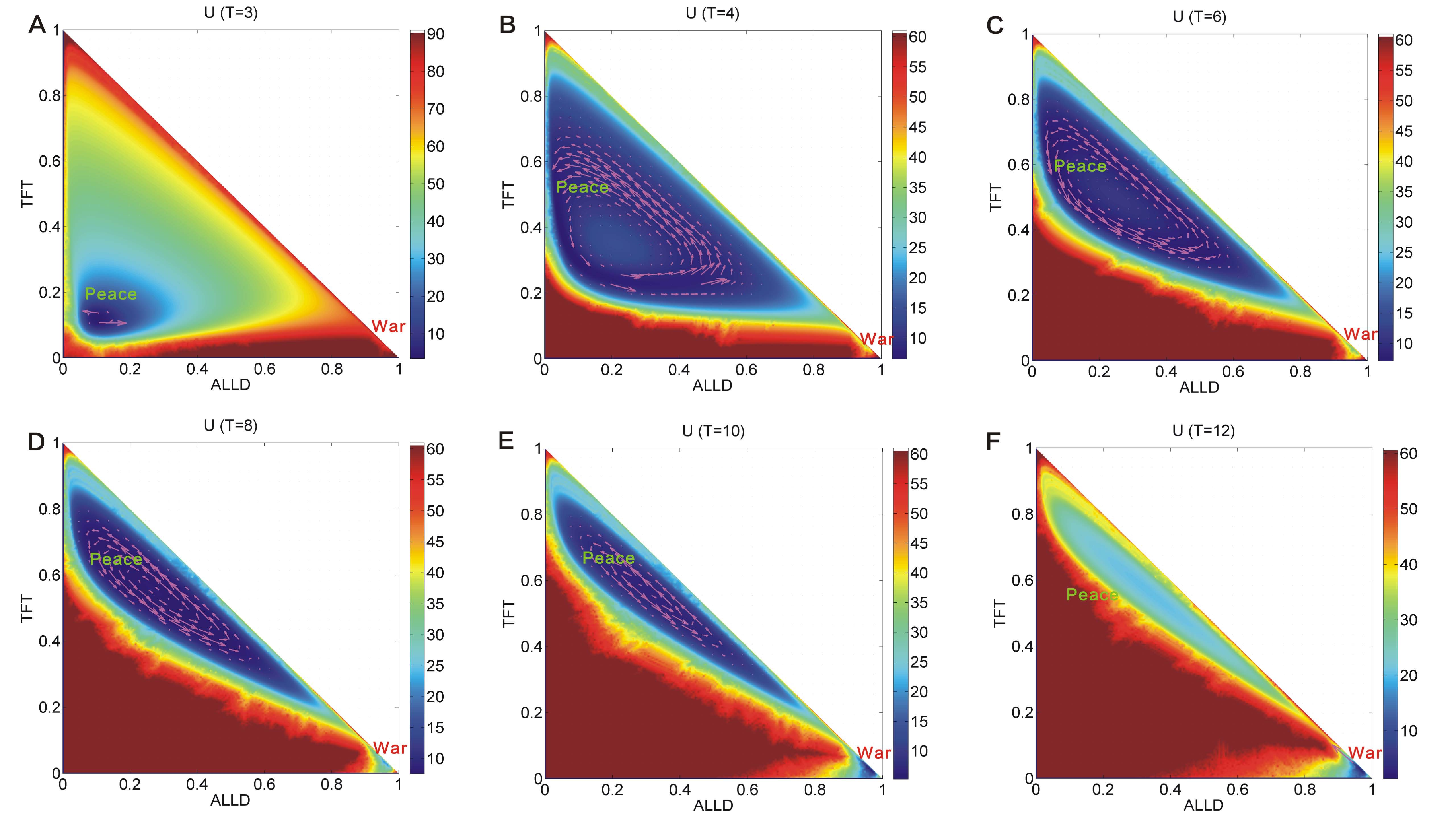}
\caption{The population landscape $U$ with different parameter $T$ and the constant parameters $\mu=0.006,c=0.22,R=3,Pu=1.0,S=0.0$.}\label{T_U}%
\end{figure*}

We show that the population landcapes $U$ and the flux change with the increasing temptation $T$ of payoff matrix (which denotes a defector acquires a temptation payoff $T$ when they encountered cooperators) as shown in Figure \ref{T_U}. The system has two stable states when temptation is very small $T=3$ where the relationship $T>R$ should hold in Figure \ref{T_U}(A). The system covers a large area of state space for limit cycle valley, a $Peace$ state and a $War$ state. We can see the basin of $Peace$ state is very stable and deeper while the $War$ state is much shallower and less stable. This illustrates majority of players choose the cooperation $ALLC$ or $TFT$ strategy which leads to more stable $Peace$ state, rather than defection strategy $ALLD$ when the temptation is less. Defection can not earn more profits. As temptation increases, the $Peace$ state is unstable and becomes a limit cycle state. The limit cycle state adopts the mixed strategy $TFT$ and $ALLD$. The limit cycle valley representing the $Peace$ state becomes shallower while the stable $War$ state becomes more stable shown in Figure \ref{T_U}(B)(C)(D). When the temptation for this game increases even further, more and more player choose defection strategy $ALLD$ to earn more profits rather than strategy $ALLC$. This is because more temptation from the defection strategy can lead to more stable $War$ state.

\subsection*{Phase diagrams, state switching, landscapes and fluxes upon cooperation reward}

Figure \ref{TRP_phase_EPR}(B) shows the phase diagram for the repeated Prisoner's Dilemma with the constant parameter $\mu=0.006,c=0.22,T=5.0,Pu=1.0,S=0.0$ and changes in parameter $R$. The element $R$ of payoff matrix denotes the reward that cooperators will acquire from the mutual aid when the cooperators encountered. The value of reward $R$ should satisfy $R>(T+S)/2$. If this relationship is not satisfied, the agreement of alternate cooperation and defection will earn more payoff than that of pure cooperation in a repeated Prisoner's Dilemma game\cite{Nowak2006Book}. There are three regions in this phase diagram. When the reward $R$ is smaller, a limit cycle and the stable $War$ state  coexists in Region III. As the reward $R$ increases, the $War$ state disappears after the saddle-node bifurcation and the limit cycle is left in Region II. As the reward $R$ increases further, the limit cycle diminishes and a stable  $Peace$ state emerges in Region I. Figure \ref{TRP_phase_EPR}(E) also shows $EPR$ versus reward $R$, which implies that the limit cycle state costs more energy while one stable state costs less energy.

\begin{figure*}[!ht]
\centering
\includegraphics[width=0.8\textwidth]{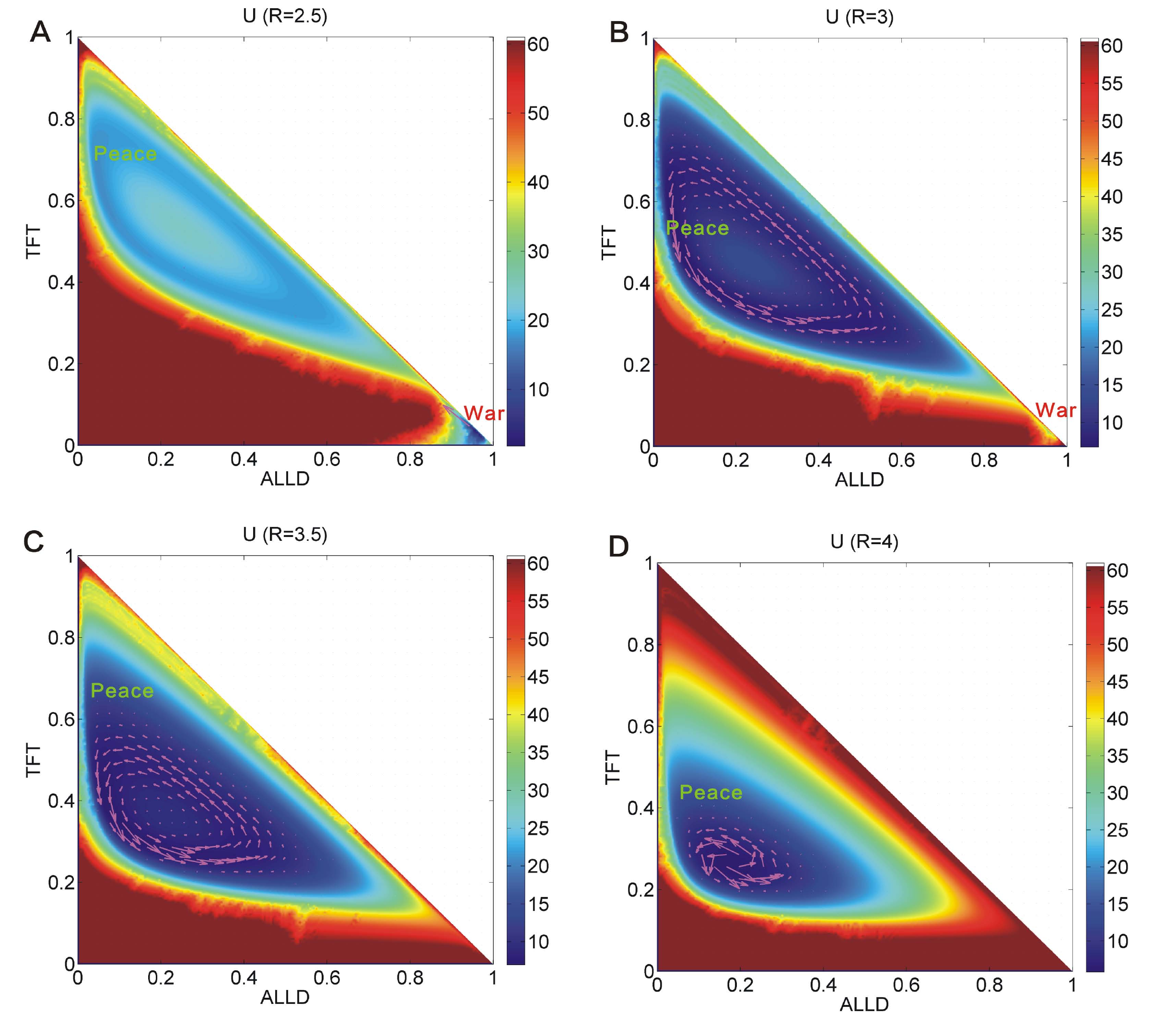}
\caption{The population landscape $U$ with different parameter $R$ and the constant parameters $\mu=0.006,c=0.22,T=5,Pu=1.0,S=0.0$.}\label{R_U}%
\end{figure*}

Figure \ref{R_U} shows the population landscape $U$ with increasing cooperation $R$. The system has one deep stable $War$ state and a shallower limit cycle ring valley for smaller parameter $R=2.5$, since the reward for cooperation strategy is much less. The majority players choose the defection strategy  leading to the $War$ state shown in Figure \ref{R_U}(A). When reward $R=3$ increases, a limit cycle valley becomes deeper and stable while the $War$ state becomes shallower and less stable as shown in Figure \ref{R_U}(B). It shows that more and more players prefer the strategy $ALLC$ rather than strategy $ALLD$ since more reward can be obtained from cooperation. As reward $R$ increases further, the limit cycle valley becomes deeper, and the $War$ state vanishes shown in Figure \ref{R_U}C. As reward $R$ increases even further, the limit cycle ring valley shrinks into a stable $Peace$ state shown in Figure \ref{R_U}D. The majority players choose cooperation strategy. It is because more reward from the cooperation strategy can lead to more stable $Peace$ state. It turns out that more reward from cooperation leads to $Peace$ with win-win outcome.

\subsection*{Phase diagrams, state switching, landscapes and fluxes upon defector punishment}

Figure \ref{TRP_phase_EPR}(C) shows the phase diagram for the repeated Prisoner's Dilemma with the constant parameter $\mu=0.006,c=0.22,T=5.0,R=3.0,S=0.0$ and changes in parameter $Pu$. The element $Pu$ of payoff matrix denotes that the defectors obtain a punishment $Pu$ when the defectors are encountered. The value of punishment $Pu$ should satisfy $R>Pu>S$ \cite{Nowak2006Book}. There are five regions in this phase diagram. When the punishment $Pu$ is smaller, only the stable $War$ state exists in Region I on the left. As the punishment $Pu$ increases, two unstable states emerge by the saddle-node bifurcation in Region IV. As the punishment $Pu$ increases further, a limit cycle emerges and coexists with the stable state $War$ in Region III; as the punishment $Pu$ increases even further, the limit cycle diminishes and a stable $Peace$ state coexists with stable $War$  state in Region V. As the punishment $Pu$ approaches to $3$, the stable $Peace$ state diminishes after the saddle-node bifurcation, and the stable $War$ state is left in Region I on the right. Figure \ref{TRP_phase_EPR}(F) also shows $EPR$ versus punishment $Pu$, which shows that the system with limit cycle dissipates more energy.

\begin{figure*}[!ht]
\centering
\includegraphics[width=0.8\textwidth]{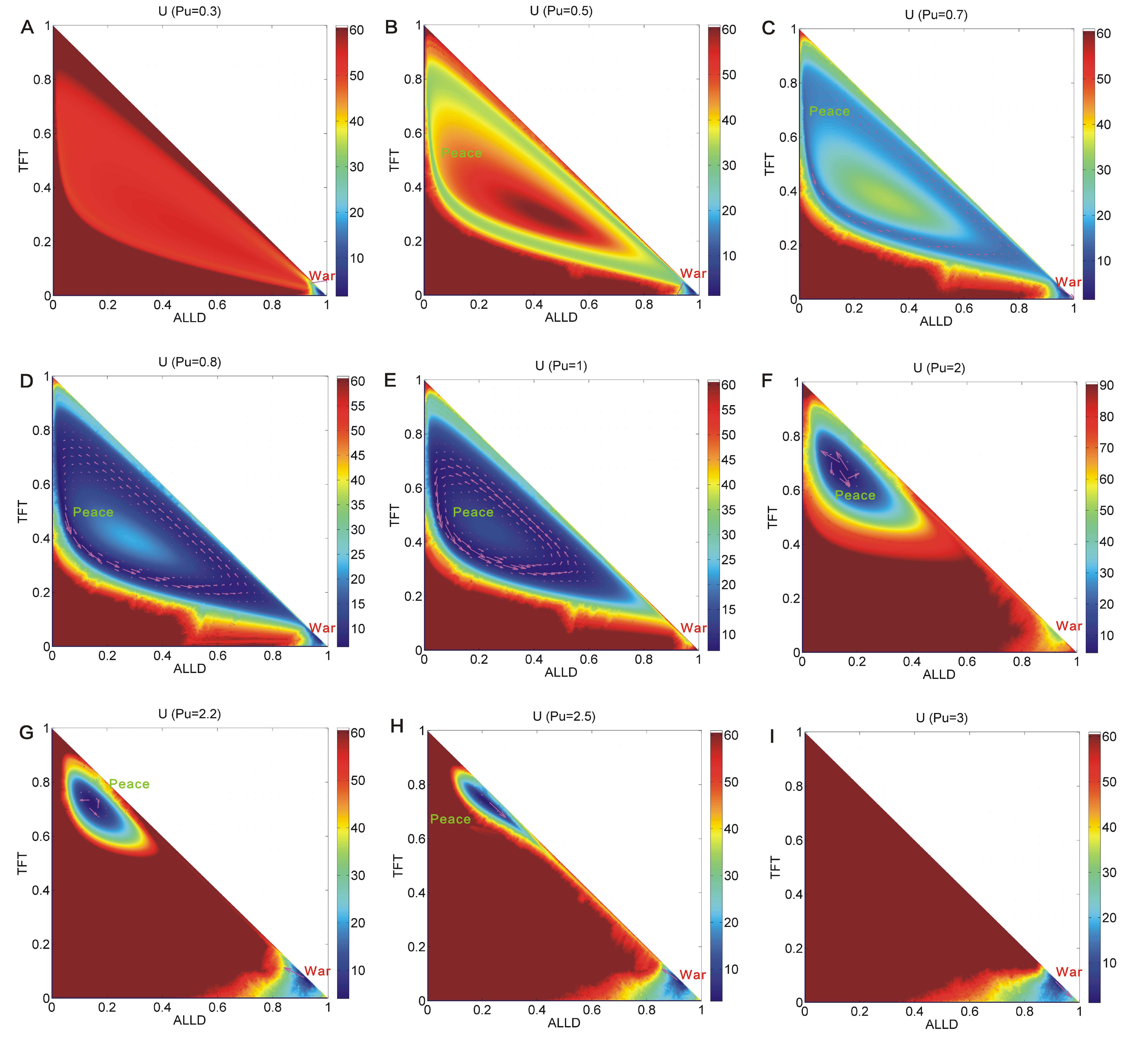}
\caption{The population landscape $U$ with different parameter $Pu$ and the constant parameters $\mu=0.006,c=0.22,T=5,R=3,S=0.0$.}\label{Pu_U}%
\end{figure*}

Figure \ref{Pu_U} shows the population landscape $U$ with increasing element $Pu$ of payoff matrix. When punishment $Pu$ is small representing that the defector can earn less from the game, almost all players choose strategy $ALLD$ shown in Figure \ref{Pu_U}(A). $ALLD$ is the only strict Nash solution and the only evolutionary stable strategy. And when the punishment ($Pu=0.3$) is small, the value of the first element in the left column of the payoff matrix for $ALLD$ is $Pu$ while the second of that for $TFT$ is $Pu-c=0.08$. Thus the fitness for $ALLD$ is much higher than that of $TFT$, and that of $ALLC$ (The sucker's payoff $S=0$). This illustrates that more and more players give up $ALLD$ since the profits of $TFT$ players are catching up with that of $ALLD$ players when they both encounter the $ALLD$ players. As punishment $Pu$ increases, a shallow limit cycle valley for parameter $Pu=0.5$ emerges accompanied with the stable $War$ state shown in Figure \ref{Pu_U}(B). When punishment $Pu$ increases further, the limit cycle valley shrinks its size but becomes deeper and more stable while the $War$ state becomes shallower and less stable. This is shown in Figure \ref{Pu_U}(C)(D)(E). Then the limit cycle ring valley shrinks to a stable $Peace$ state where more players choose $TFT$ strategy as shown in Figure \ref{Pu_U}(F)(G)(H). It shows that the $Peace$ state becomes more stable first and then loses its stability while the $War$ state becomes more stable as punishment $Pu$ increases. It demonstrates that the punishment $Pu$ has an optimal value to lead to a more stable $Peace$ state. At last, when the value of punishment $Pu$ is approaching to the value of reward $R$, the stable $Peace$ state diminishes and the stable $War$ state is left. This shows that when the punishment $Pu$ and the reward $R$ are almost the same, $ALLD$ is dominant than that of $TFT$ since $ALLD$ is the only strict Nash solution and the evolutionary stable state. These results show that the strategy $ALLD$ is a dominant strategy and has an advantage for selection.

We explored the free energy versus the parameters of the repeated Prisoner's Dilemma model in Figure \ref{cm_phase_EPR}(E) (the free energy versus cost $c$), Figure \ref{cm_phase_EPR}(F) (the free energy versus the mutation rate $\mu$), Figure \ref{TRP_phase_EPR}(G) (the free energy versus temptation $T$), Figure \ref{TRP_phase_EPR}(H) (the free energy versus reward $R$), Figure \ref{TRP_phase_EPR}(I) (the free energy versus punishment $Pu$). We can see this five free energy profiles have some similarity that each has the opposite tendency with that of the corresponding $EPR$. These free energies link to the different phases and phase transitions. The first derivative of the free energies have discontinuous at the transition points from a stable state to a limit cycle oscillation state and vice versa. It implies non-equilibrium thermodynamic phase transition has certain similarities as that of the equilibrium statistical mechanics. Free energy profiles can manifest the phase transitions and be used to explore the global stability and robustness of the game system.

\subsection*{Kinetic speed and optimal paths of switching between two stable states}

We explored the escape time for the evolutionary game system. We used the following formula to solve the escape time $\tau$: $ \mathbf{F} \cdot \nabla \tau + D \cdot \nabla^2 \tau =-1$\cite{VanKampen}. The escape time can be viewed the average time spent for a system from one state to another \cite{Zhang2012JCP}. We can set the mean first passage time (MFPT) $\tau_{pw}$ representing the MFPT from  $Peace$ state to $War$ state and $\tau_{wp}$ representing the MFPT from $War$ state to $Peace$ state.

We also explored the kinetic paths of the repeated Prisoner's Dilemma model. We can obtain the relative probabilities of each path using statistics and the quantification of the path weights. Path integral weights can be calculated by the action, which is analogous to the classical mechanical systems. The dominant paths with the
largest weights can be viewed the major pathways. The path integral formulate is shown as\cite{Wang2010JCP,Zhang2012JCP}:

\begin{eqnarray}
P({\bf x_{final}},t,{\bf x_{initial}},0)&=&\int D {\bf x}
Exp[-\int dt ( \frac{1}{2} \nabla \cdot {\bf F} ( {\bf x} )
\nonumber \\&& + \frac{1}{4}  ( d {\bf x}/{dt} - {\bf F} ({\bf x}
) ) \cdot \frac{1}{ {\bf D} ( {\bf x} ) } \cdot ( d {\bf x}/{dt}
- {\bf F} ({\bf x} ) ) ) ]\nonumber\\
&=&\int D {\bf x}Exp[-A(\bf x)]
\end{eqnarray}

The above probability describes the chance of starting at the point $\bf x_{initial}$ at initial time and ending at the point $\bf x_{final} $ at the final time. The probability is the result of the sum of the weights from all possible paths. $A(\bf x)$ is the action for each path. Each weight is exponentially related to the action which has two contributions, one from the stochastic equation of motion for dynamics and the other is from the variable transformation from the stochastic force to the system variable. Not all the paths contribute equally to the weight. Due to the exponential nature, the optimal path is exponentially larger in weight than the suboptimal ones. Therefore, we can identify the optimal path with the most probable weight.

\begin{figure*}[!ht]
\centering
\includegraphics[width=0.8\textwidth]{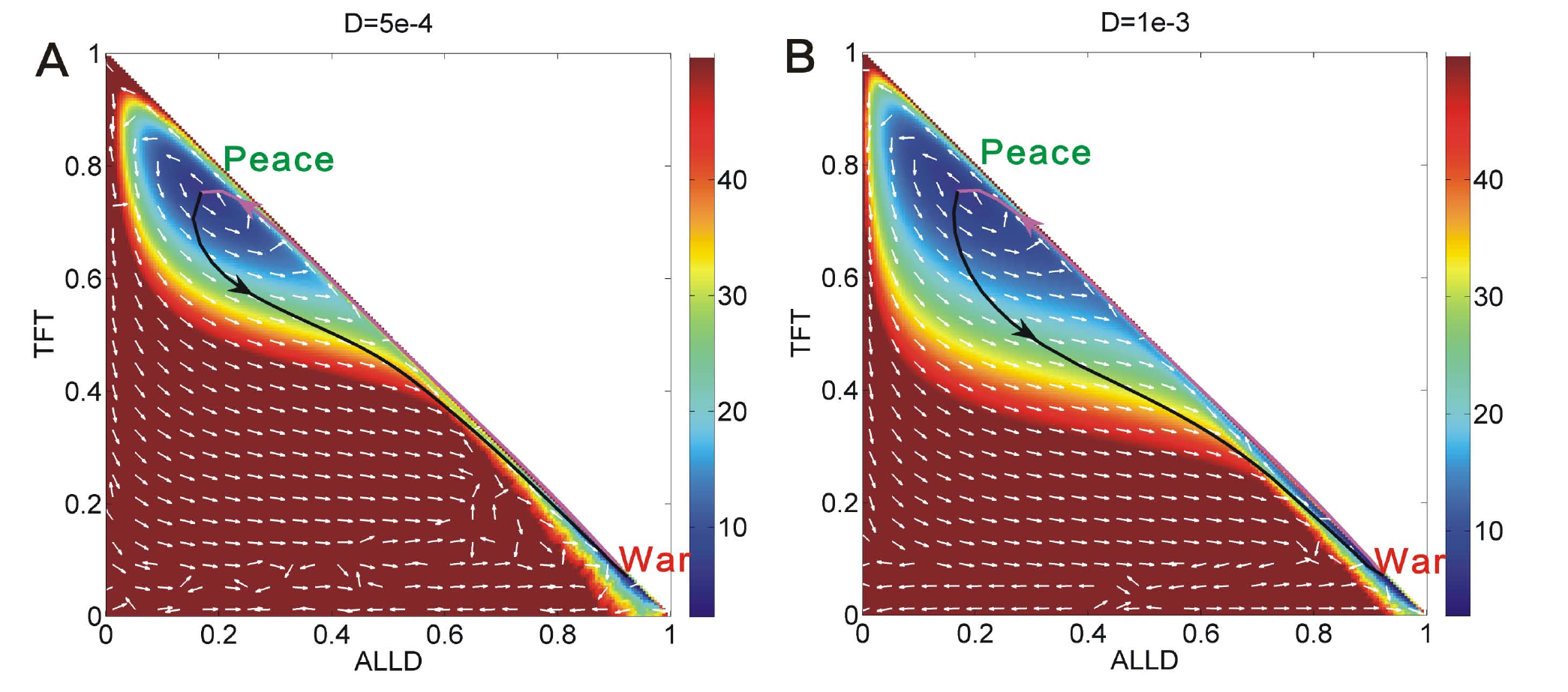}
\caption{The pathways on the population potential landscape $U$ with different diffusion coefficient $D$ and $\mu=0.006,c=0.22,T=5,R=3,Pu=2.4,S=0.0$. The purple lines represent the paths from the $War$ state to $Peace$ state. The black lines represent the paths from the $Peace$ state to $War$ state. The white arrows represent the probability fluxes.}\label{Pu_pathD}%
\end{figure*}

We studied the repeated Prisoner's Dilemma model with the parameters nearby the set of $\mu=0.006,c=0.22,T=5,R=3,Pu=2.4,S=0.0$ which has two stable state $Peace$ and $War$. Figure \ref{Pu_pathD} shows the optimal paths on the population landscape $U$ with different diffusion coefficient $D$ and $\mu=0.006,c=0.22,T=5,R=3,Pu=2.4,S=0.0$. We can see there are two stable states :$War$ and $Peace$ on the population landscapes. The purple lines represent the paths from the $War$ state to $Peace$ state. The black lines represent the paths from the $Peace$ state to $War$ state. The white arrows represent the probability fluxes which guide the paths apart from the steepest descent path from the landscape. Therefore, the optimal path from $War$ to $Peace$ states and the optimal path from $Peace$ to $War$ states are apart from each other. Under more fluctuations (bigger diffusion coefficient $D$ shown in Figure \ref{Pu_pathD}(B)), the two paths are further apart from each other due to larger probability fluxes. We can see the purple lines and black lines are irreversible in both two sub figures. These lines are apart from each other due to the non-zero flux. The optimal paths are deviated from the naively expected steepest descent paths from potential landscape. We can clearly see the fluxes have spiral shapes which show the property of non-equilibrium system.

\begin{figure*}[!ht]
\centering
\includegraphics[width=0.8\textwidth]{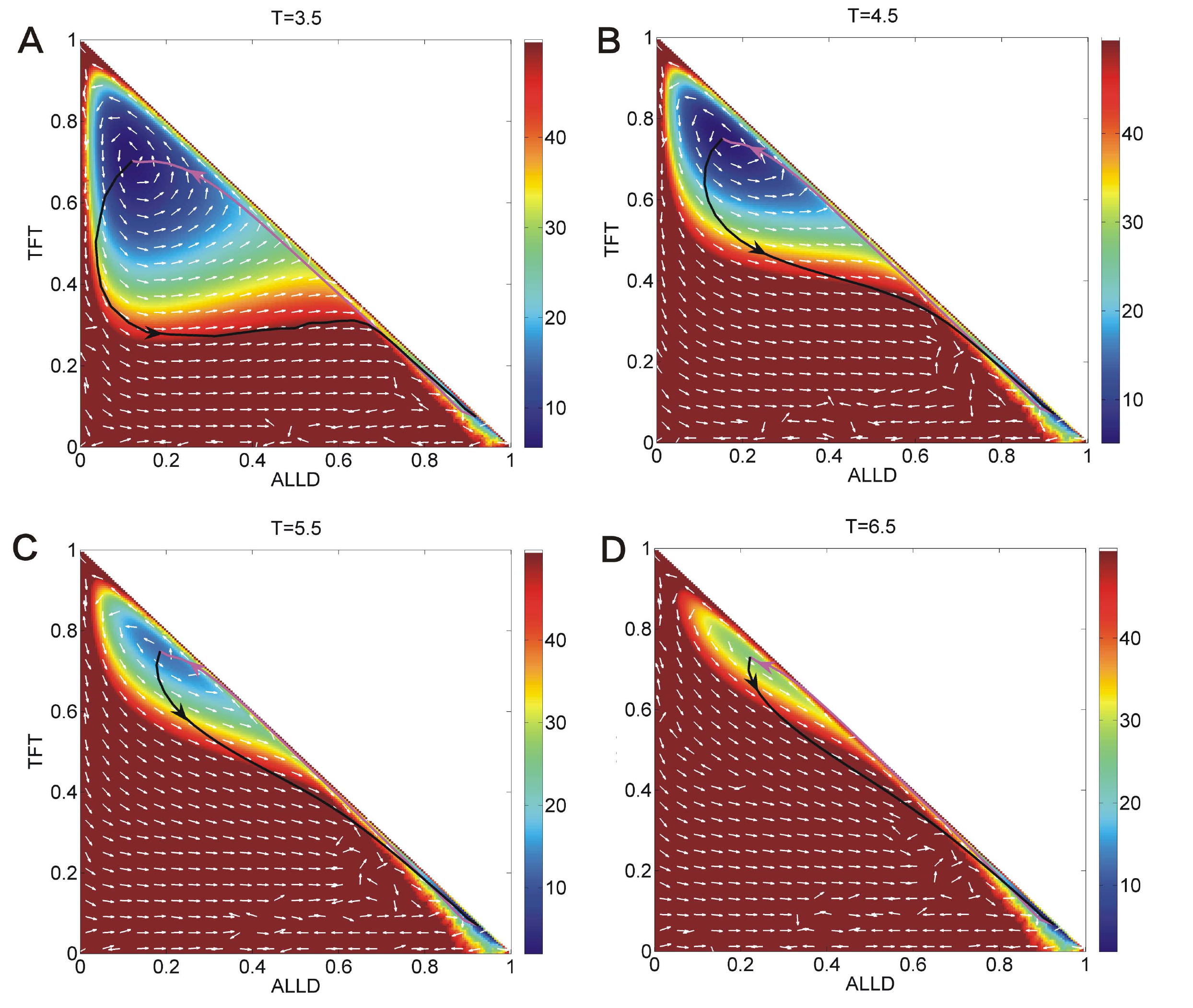}
\caption{The population landscape $U$ with different parameter $T$ and the constant parameters $\mu=0.006,c=0.22,R=3,Pu=2.4,S=0.0$. The purple lines represent the paths from the $War$ state to $Peace$ state. The black lines represent the paths from the $Peace$ state to $War$ state. The white arrows represent the probability fluxes.}\label{T_path}%
\end{figure*}

\begin{figure*}[!ht]
\centering
\includegraphics[width=1.0\textwidth]{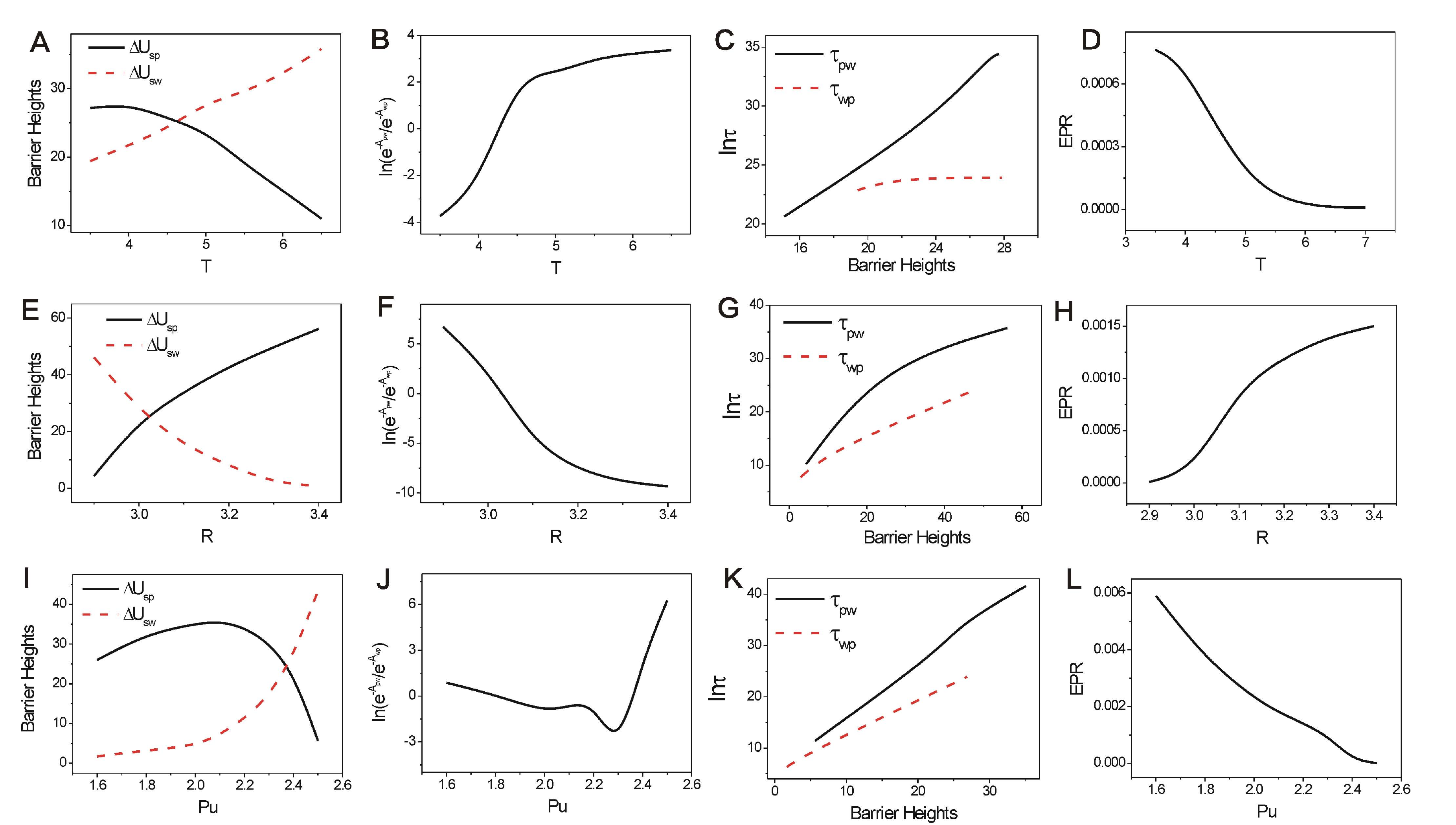}
\caption{A: the barrier heights versus $T$, B: the logarithm of the ratio of the probability of dominant path from state $Peace$ to state $War$ and the probability of dominant path from state $War$ to state $Peace$ versus $T$, C: the logarithm of the escape time $MFPT$ versus $T$, D: entropy production rate versus $T$, E: the barrier heights versus $R$, F: the logarithm of the ratio of the probability of dominant path from state $Peace$ to state $War$ and the probability of dominant path from state $War$ to state $Peace$ versus $R$, G: the logarithm of the escape time $MFPT$ versus $R$, H: entropy production rate versus $R$, I: the barrier heights versus $Pu$, J: the logarithm of the ratio of the probability of dominant path from state $Peace$ to state $War$ and the probability of dominant path from state $War$ to state $Peace$ versus $Pu$, K: the logarithm of the escape time $MFPT$ versus $Pu$, L: entropy production rate versus $Pu$.}\label{TRP_Ba_EPR}%
\end{figure*}

\begin{figure*}[!ht]
\centering
\includegraphics[width=0.8\textwidth]{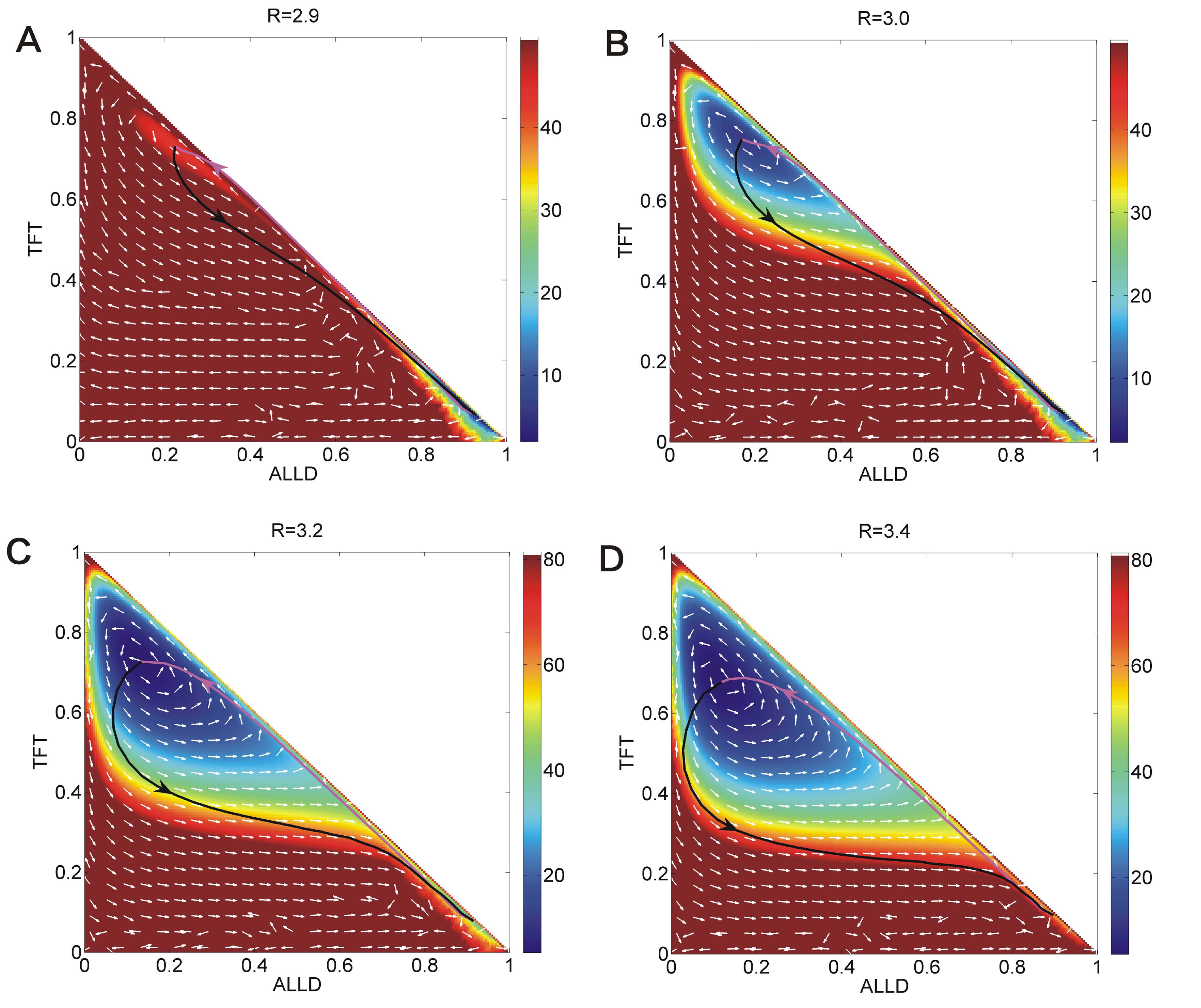}
\caption{The population landscape $U$ with different parameter $R$ and the constant parameters $\mu=0.006,c=0.22,T=5,Pu=2.4,S=0.0$. The purple lines represent the paths from the $War$ state to $Peace$ state. The black lines represent the paths from the $Peace$ state to $War$ state. The white arrows represent the probability fluxes.}\label{R_path}%
\end{figure*}

\begin{figure*}[!ht]
\centering
\includegraphics[width=0.8\textwidth]{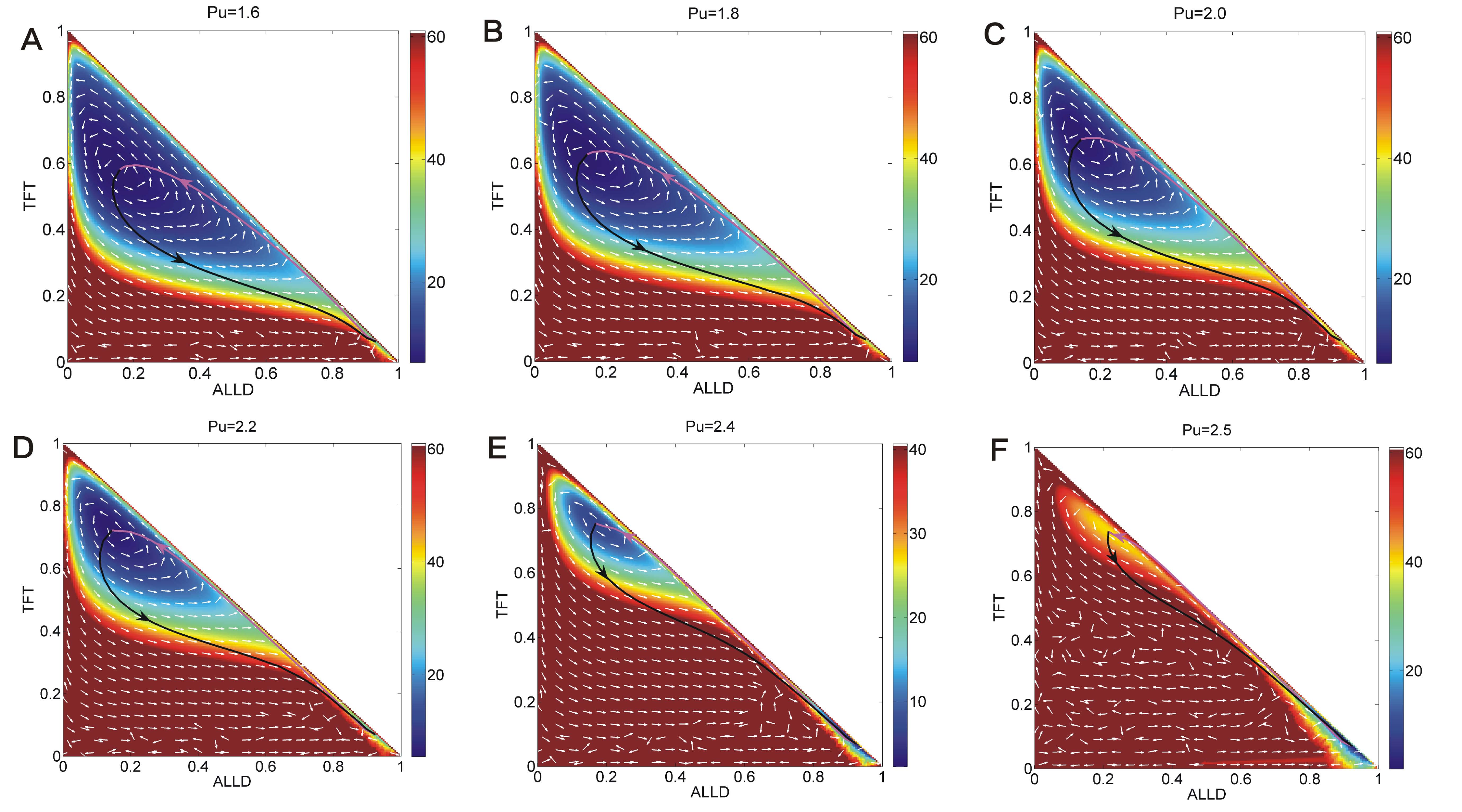}
\caption{The population landscape $U$ with different parameter $Pu$ and the constant parameters $\mu=0.006,c=0.22,T=5,R=3,S=0.0$. The purple lines represent the paths from the $War$ state to $Peace$ state. The black lines represent the paths from the $Peace$ state to $War$ state. The white arrows represent the probability fluxes.}\label{Pu_path}%
\end{figure*}

We show the population landscape $U$ and the paths with different parameter temptation $T$ and the constant parameters $\mu=0.006,c=0.22,R=3,Pu=2.4,S=0.0$ in Figure \ref{T_path}. The purple lines represent the paths from the $War$ state to $Peace$ state. The black lines represent the paths from the $Peace$ state to $War$ state. The white arrows represent the probability fluxes. The paths are deviated from the naively expected steepest descent paths from potential landscape, and they are irreversible. The two paths become more closer from each other as $T$ increases. Figure \ref{TRP_Ba_EPR}(A) shows the barrier heights versus temptation $T$. We set $\Delta U_{sp}=U_{saddle}-U_{Peace}$ as the barrier height between state $Peace$ and the saddle point and $\Delta U_{sw}=U_{saddle}-U_{War}$ as the barrier height between state $War$ and the saddle point, where $U_{saddle}$ represents the value of landscape $U$ at the saddle point between state $Peace$ and $War$, $U_{Peace}$ represents the minimum value of landscape $U$ at state $Peace$, $U_{War}$ represents the minimum value of landscape $U$ at state $War$. We can see that the barrier height $\Delta U_{sw}$ increases while the barrier height $\Delta U_{sp}$ decreases. It means the state $Peace$ loses stability as the state $War$ becomes more stable as temptation $T$ increases. It denotes temptation guides more players to choose strategy $ALLD$. The path weight denotes the probability of each route. The path probability can be obtained by the action $A(x)$ for Peace and War switching. We labeled $A_{wp}$ as the action of the dominant path from state $War$ to state $Peace$, and $A_{pw}$ as the action of the dominant path from state $Peace$ to state $War$. Figure \ref{TRP_Ba_EPR}(B)showed the logarithm of $Peace$ to $War$ path probability divided that of $War$ to $Peace$ path increases as temptation $T$ increases. It represents that the path from state $Peace$ to state $War$ has more probability than that of state $War$ to state $Peace$.

We show the population landscape $U$ and the paths with different parameter reward $R$ and the constant parameters $\mu=0.006,c=0.22,T=5,Pu=2.4,S=0.0$ in Figure \ref{R_path}. The paths are not following the steepest descent paths from potential landscape. The two paths depart far away from each other as $R$ increases. Figure \ref{TRP_Ba_EPR}(E) shows the barrier height $\Delta U_{sw}$ decreases while the barrier height $\Delta U_{sp}$ increases as reward $R$ increases. Figure \ref{TRP_Ba_EPR}(F)showed the logarithm of $Peace$ to $War$ path
probability divided that of $War$ to $Peace$ path decreases as reward $R$ decreases. It represents that as the reward increases, more and more players choose strategy $ALLC$ which guides the $Peace$ state more state while the $War$ state less stable, and then the path from state $War$ to state $Peace$ has more probability then opposite path.

We show the population landscape $U$ and the paths with different parameter punishment $Pu$ and the constant parameters $\mu=0.006,c=0.22,T=5,R=3,S=0.0$ in Figure \ref{Pu_path}. The two paths are not following the steepest descent paths from potential landscape and come closer from each other as $Pu$ increases. Figure \ref{TRP_Ba_EPR}(I)shows the barrier height $\Delta U_{sw}$ increases while the barrier height $\Delta U_{sp}$ increases first and then decreases as punishment $Pu$ increases. Figure \ref{TRP_Ba_EPR}(J)showed the logarithm of $Peace$ to $War$ path probability divided that of $War$ to $Peace$ path decreases first and then increases as punishment $Pu$ increases. It represents that as the punishment increases, more players choose strategy $ALLC$ and $ALLD$ which guides the $Peace$ state and $War$ state both becoming more state. As the punishment grows bigger, the $War$ state becomes more stable than that of state $Peace$. It denotes that the value of punishment $Pu$ has an optimal value to make more stable $Peace$ state.

Figure \ref{TRP_Ba_EPR}(C)(G)(K)shows the logarithm of the MFPT increases as their corresponding barrier heights increase corresponding to Figure \ref{TRP_Ba_EPR}(A)(E)(I). The logarithm of the MFPT and barrier heights have the positive relationship as following: $\tau=exp(\Delta {Ba})$. The average kinetic speeds of state transition along the according paths can be measured by the $1/\tau$. As the barrier height becomes higher, the escape time becomes longer and the kinetic speed becomes slower. Therefore, the state is more stable with higher barrier height. It is more difficult to switch from one basin of attraction to another with higher barriers between two stable states. It takes more time to transition and the kinetic speed is slower. The $MFPT$, kinetic speed and barrier height can supply the measurements of quantifying the stability of game theory systems. And they can give more information about the dynamics of a game theory system. It also can interpret the mechanisms of the state transition from the game theory systems.

Figure \ref{TRP_Ba_EPR}(D)(H)(L)shows the entropy production rate versus parameters $T,R,Pu$ among the phase rang of two stable state. We explored the linear stability of the repeated prisoners' dilemma. The eigenvalues of state $Peace$ have negative real parts and two opposite number imaginary parts. It denotes that state $Peace$ is a stable focus which oscillates and spiral to its destiny. Stable focus will transit to an unstable focus which is a limit cycle. The eigenvalues of state $War$ have negative real parts and no imaginary parts. It denotes that state $War$ is a stable node. $EPR$ decreased as temptation $T$ increases, while $Peace$ state loses its stability and $War$ state becomes more stability shown in Figure \ref{TRP_Ba_EPR}(D). $EPR$ increased as reward $R$ increases, while $War$ state loses its stability and $Peace$ state becomes more stability shown in Figure \ref{TRP_Ba_EPR}(H). $EPR$ decreased as punishment $Pu$ increases, while $Peace$ state loses its stability and $War$ state becomes more stability shown in Figure \ref{TRP_Ba_EPR}(L). We can clear see that the system with dominant $Peace$ state will cost more entropy production rate $EPR$ than that with dominant $War$ state. This is because stable focus will cost more energy. It represents that keeping the peace will consume more energy.

\newpage
\newpage
\newpage
\newpage
\newpage
\section*{Discussions and Conclusion}

Global stability and the underlying mechanism of the dynamics are crucial for understanding the nature of the game theory. Foster and Young presented the analysis of stochastic perturbation of evolutionary game dynamics, defining the stability for a stochastic dynamics. It is viewed as capturing the long-run stability of the stochastic evolutionary game dynamics rather than the evolutionary stable strategy and the Nash equilibrium\cite{Szabo2007PR,Foster1990TPB}. They also introduced the idea of a potential function which can be used to compute the stochastically stable set. However, their method can only obtain the potential function in one dimension and often in equilibrium. In reality, the systems are often more complex. The evolutionary game dynamics are general in non-equilibrium. It is difficult to obtain the analytical potential functions to capture the stability of the evolutionary game systems in higher dimensions, since a pure gradient of potential landscape cannot be obtained for general evolutionary game dynamics.

Many researchers tried to explore the stability of the evolutionary game systems. Some chose the simulations of the trajectories under fluctuations\cite{Foster1990TPB, Imhof2005Evolutionary}. The stability of the repeated Prisoner's Dilemma game can be explored by the Lyapunov function. But the Lyapunov function cannot be found easily. In this work we develop the landscape and flux theory for quantifying the population and intrinsic landscape. The intrinsic landscape $\phi_0$ has a Lyapunov property which can be used to explore the global stability of the game theory systems. We obtained the numerical Lyapunov function $\phi_0$ by solving the Hamilton-Jacobi equation and the population potential $U$ from the Fokker-Plank diffusion equation. Thus we can explore the global
stability of the game theory system by the intrinsic landscape $\phi_0$ and the population potential landscape $U$. The repeated Prisoner's Dilemma game system is a non-equilibrium system. The underlying dynamics is determined by both the force from the gradient of the landscape and the force from the probability flux which breaks the detailed balance. This provides a new view to explore the game theory dynamics. In conventional evolutionary game dynamics, the flux is not considered. Here we consider both the landscape and flux to study the evolutionary game dynamics.

The barrier height and the entropy production rate can quantify the global stability of the non-equilibrium repeated Prisoner's Dilemma game. The irreversible paths are obtained by the path integral method. The optimal paths are not along the gradient of the landscape due to the non-zero probability flux. We also quantified the mean first passage time which can measure the kinetic speed of the dynamics of switching from one state to another.

We have found that when the cost for $TFT$ which reduces the values of element of $TFT$ in payoff matrix is small, and thus the values of payoff elements for $TFT$ are large, the system approaches to $Peace$ state easily. As the cost $c$ increases further, the system will go to the $War$ state since the profits from the $TFT$ strategy is much less. We have also found that when $c$ is small, high mutation rate will lead to $Peace$ state far from $War$ state.

When $c$ is moderate, high mutation rate will lead to a mixed strategy state which has almost the same probability of these three strategies. It prefers the system far from $War$ state. However, as the cost $c$ is larger, the system will fall into $War$ state either with low mutation rate or high mutation rate.

We have also found that  moderate intensity of punishment for defection strategy (moderate value of parameter $Pu$) decreases the stability of $War$ state. More reward for cooperation strategy (high value of parameter $R$) prefers the $Peace$ state. More temptation for defector from cooperator will prefer $War$ state to earn more profits using defection strategy. Thus choosing a moderate intensity of punishment for defection strategy and increasing the intensity of reward for cooperation strategy will avoid the lasting $War$ state, and favor the  long lasting $Peace$ state.

The oscillations between $Peace$ and $War$ can be explained as: when $Peace$ state sustains for a long time, the population increases and the resource relatively reduces. In order to survive, the populations fight for the resources, to get better livings. After a long war, the populations do not engage in the production, livelihood, and fall into a long-term state of tension. When the $Peace$ state emerges, the population grows. The oscillations will circulate.

We provided the integral pathway method to obtain the paths between each two stable state. The paths between two states $Peace$ and $War$ are irreversible due to the non-zero flux which is the characteristic for non-equilibrium system. The probability of each path also can give us the information about stability. More stable state has less probability of the corresponding path to exit from its attraction. It spent more time to exit from more stable state with higher barrier heights as $MFPT$ shown, thus the speed of transition between stable states are slower. We also obtained that the system with dominant stable focus $Peace$ has more $EPR$. It means that keeping peace will cost more energy. This method supplied a chance for people to explore the properties of the paths and the kinetics speeds with their according barrier heights for game theory systems.

Since the stochastic game theory dynamics is more difficult to explore analytically, we provide a potential-flux framework to explore and quantify the stochastic game theory dynamics. The investigations of the global stability are essential for understanding the nature and the underlying mechanisms of the game theory dynamics. We show this in an example of repeated Prisoner's Dilemma game system. This can help the further understanding of the game theory for the real world.

\section*{Acknowledgments}
This project is supported by Natural Science Foundation of China No.11305176,91430217, MOST, China, Grant No.2016YFA0203200, and Jilin Province Youth Foundation of China Grant No.20150520082JH, China. JW thanks the support from grant no. NSF-PHY 76066.


\end{document}